\documentclass[dvipdfm,epsf]{iopart}
\usepackage{iopams,amsfonts,rotating,graphicx,fancybox,fancyhdr,bm,eucal,dsfont,appendix}
\usepackage{datetime}

\eqnobysec

\settimeformat{hhmmsstime}
\ddmmyyyydate

\begin{document}

\title[Integrable Aspects of Universal Quantum Transport in Chaotic Cavities]
        {Integrable Aspects of Universal Quantum Transport in Chaotic Cavities\;\footnote[3]{To Appear in the Special Issue: Painlev\'e Equations -- Part II (Constructive Approximation, 2014)}}

\author{Eugene Kanzieper}

\address{Department of Applied Mathematics, Holon Institute of Technology, Holon 5810201, Israel
    } \eads{\mailto{eugene.kanzieper@hit.ac.il}}

\begin{abstract}
The Painlev\'e transcendents discovered at the turn of the XX century by pure mathematical reasoning, have later made their surprising appearance -- much in the way of Wigner's ``miracle of appropriateness'' -- in various problems of theoretical physics. The notable examples include the two-dimensional Ising model, one-dimensional impenetrable Bose gas, corner and polynuclear growth models, one dimensional directed polymers, string theory, two dimensional quantum gravity, and spectral distributions of random matrices. In the present contribution, ideas of integrability are utilized to advocate emergence of an one-dimensional Toda Lattice and the fifth Painlev\'e transcendent in the paradigmatic problem of conductance fluctuations in quantum chaotic cavities coupled to the external world via ballistic point contacts. Specifically, the cumulants of the Landauer conductance of a cavity with broken time-reversal symmetry are proven to be furnished by the coefficients of a Taylor-expanded Painlev\'e V function. Further, the relevance of the fifth Painlev\'e transcendent for a closely related problem of sample-to-sample fluctuations of the noise power is discussed. Finally, it is demonstrated that inclusion of tunneling effects inherent in realistic point contacts does not destroy the integrability: in this case, conductance fluctuations are shown to be governed by a two-dimensional Toda Lattice.
\end{abstract}

\newpage
\tableofcontents

\newpage
\section{Introduction} \label{intro}
The low temperature electronic conduction through a cavity exhibiting chaotic classical dynamics is governed by quantum
phase-coherence effects (Imry 2002). In chaotic cavities with sufficiently large capacitance, when
electron-electron interactions are negligible, the most
comprehensive theoretical framework by which the phase coherent
electron transport can be explored is provided by the scattering
${\bm {\mathcal S}}$-matrix approach pioneered by Landauer (Landauer 1957, Fisher and Lee 1981, B\"uttiker 1990). There exist two different, though mutually
overlapping, scattering-matrix descriptions (Lewenkopf and Weidenm\"uller 1991) of
quantum transport.

A semiclassical formulation \footnote[4]{The supersymmetry approach due to Efetov (1997) is yet another tool for a non-perturbative description of quantum transport through tiny chaotic/disordered structures.} (Richter 2000) of the ${\bm {\mathcal
S}}$-matrix approach is tailor-made to the analysis of
energy-averaged charge conduction through an individual
cavity. (Energy averaging is performed over such a small energy window near the Fermi
energy that keeps the classical dynamics essentially unchanged.) Representing quantum transport observables (such as
conductance, shot-noise power, transferred charge etc.) in terms of
classical trajectories connecting the leads attached to a cavity,
the semiclassical approach efficiently accounts for system-specific features of the quantum transport (Aleiner and Larkin 1996, 1997; Agam, Aleiner and Larkin 2000; Adagideli 2003;
 Brouwer and Rahav 2006; Whitney and Jacquod 2006; Brouwer 2007). Besides, it also covers the long-time scale {\it universal transport regime} (Richter and Sieber 2002, Heusler et al 2006, Braun et al 2006, M\"uller et al 2007) emerging in the limit $\tau_{\rm
D} \gg \tau_{\rm E}$, where  $\tau_{\rm D}$ is the average electron
dwell time and $\tau_{\rm E}$ is the Ehrenfest time (the time scale
where quantum effects set in). The Ehrenfest time $\tau_{\rm E}\simeq \lambda^{-1} \log (W/\lambda_F)$
    is determined by the Lyapunov exponent $\lambda$ of
    chaotic classical dynamics, the Fermi wavelength $\lambda_F$, and the lead widths $W$.
    The mean dwell time is estimated as $\tau_{\rm D}\simeq A/(W v_F)$, where $A$ is
    the area of the cavity and $v_F$ is the Fermi velocity.

The {\it universal} regime can alternatively be
studied within a stochastic approach based on a
random matrix description of electron dynamics in a
cavity (Brouwer 1995, Beenakker 1997, Alhassid 2000). Modelling a single electron Hamiltonian by an $M \times M$
random matrix ${\bm {\mathcal H}}$ of proper symmetry, the stochastic
approach starts with the Hamiltonian $H_{\rm tot}$ of the total
system comprised by the cavity and the leads:
\begin{eqnarray}
\label{ham}
    H_{\rm tot} = \sum_{k,\ell=1}^M {\bm \psi}_k^\dagger {\mathcal H}_{k\ell}
    {\bm \psi}_\ell &+&  \sum_{\alpha=1}^{N_{\rm L}+N_{\rm R}}
    {\bm \chi}_{\alpha}^\dagger \varepsilon_F {\bm \chi}_{\alpha}\nonumber \\
    &+& \sum_{k=1}^M \sum_{\alpha=1}^{N_{\rm L}+N_{\rm R}} \left(
    {\bm \psi}_k^\dagger
    {\mathcal W}_{k \alpha} {\bm \chi}_{\alpha} +
    {\bm \chi}_{\alpha}^\dagger {\mathcal W}_{k \alpha}^* {\bm \psi}_k
    \right).
\end{eqnarray}
Here, ${\bm \psi}_k$ and ${\bm \chi}_\alpha$ are the annihilation
operators of electrons in the cavity and in the leads, respectively.
Indices $k$ and $\ell$ enumerate electron states in the cavity:
$1\le k,\ell \le M$, with $M\rightarrow \infty$. Index $\alpha$
counts propagating modes in the left ($1\le \alpha \le N_{\rm L}$)
and the right ($N_{\rm L}+1 \le \alpha \le N_{\rm R}$) lead. The
$M\times N$ matrix ${\bm {\mathcal W}}$ describes the coupling of electron
states with the Fermi energy ${\varepsilon_F}$ in the cavity to
those in the leads; $N=N_{\rm L}+N_{\rm R}$ is the total number of
propagating modes (channels). Since in Landauer-type theories [see, e.g., a recent review by Lesovik and Sadovskyy (2011)] the
transport observables are expressed in terms of the $N\times N$
scattering matrix
\begin{eqnarray}
\label{sm}
    {\bm {\mathcal S}}(\varepsilon_F) = \mathds{1}_N - 2i \pi  {\bm {\mathcal W}}^\dagger
    ({\varepsilon}_F \mathds{1}_M - {\bm {\mathcal H}} + i \pi {\bm {\mathcal W}}
    {\bm {\mathcal W}}^\dagger)^{-1} {\bm {\mathcal W}},
\end{eqnarray}
the knowledge of its distribution is central to the stochastic
approach.

For random matrices ${\bm {\mathcal H}}$ drawn from rotationally
invariant Gaussian ensembles (Mehta 2004, Forrester 2010), the distribution of
${\bm {\mathcal S}}(\varepsilon_F)$ is described by the
Poisson kernel (Hua 1963; Brouwer 1995; Mello and Baranger 1999)
\begin{eqnarray}
\label{pk}
    P({\bm {\mathcal S}}) \propto \left[
    {\rm det}\left(
        \mathds{1}_N -  {\bm {\mathcal S}_0} {\bm {\mathcal S}}^\dagger\right)
        {\rm det}\left(
        \mathds{1}_N -   {\bm {\mathcal S}} {\bm {\mathcal S}}_0^\dagger\right)
    \right]^{ \beta/2 -1  -  \beta N/2}.
\end{eqnarray}
Here, $\beta$ is the Dyson index accommodating system
symmetries \footnote[3]{We remind that $\beta=1$ corresponds to systems with preserved time-reversal and spin-orbit symmetry, $\beta=2$ refers to systems with broken time-reversal symmetry, and $\beta=4$ if the time-reversal symmetry is preserved but the spin-rotational symmetry is broken by spin-orbit scattering.} ($\beta=1,\, 2, {\rm and} \,\, 4$) whilst ${\bm {\mathcal S}}_0$ is the average
scattering matrix (Beenakker 1997),
\begin{eqnarray}\label{s0-1}
    {\bm {\mathcal S}}_0 = \frac{M \Delta {\mathds 1}_N - \pi^2 {\bm {\mathcal W}}^\dagger {\bm {\mathcal W}}}
    {M \Delta {\mathds 1}_N + \pi^2 {\bm {\mathcal W}}^\dagger {\bm {\mathcal W}}}, \qquad M\gg 1,
\end{eqnarray}
whose eigenvalues characterise couplings between the cavity and the leads in
terms of tunnel probabilities $\Gamma_j$ of $j$-th mode
in the leads ($1\le j \le N$),
\begin{equation} \label{gj}
    {\bm {\mathcal S}}_0 = {\bm V}^\dagger \,{\rm diag}
    (\sqrt{1-\Gamma_j}
    )\, {\bm V}.
\end{equation}
This relation becomes evident after one realizes that the average scattering matrix is proportional to a reflection matrix of tunnel barriers (Beenakker 1997).
Here, $\Delta$ is the mean level spacing; the matrix ${\bm V}$ is ${\bm V}\in G(N)/G(N_{\rm
L}) \times G(N_{\rm R})$ where $G$ stands for orthogonal
($\beta=1$), unitary ($\beta=2$) or symplectic ($\beta=4$) group. The celebrated result Eq.~(\ref{pk}), that can be viewed as a generalisation of the three Dyson circular ensembles,
    was alternatively derived through a phenomenological information-theoretic
    approach reviewed by Mello and Baranger (1999).

Statistical information accommodated in the Poisson kernel is too detailed to make a nonperturbative description of transport observables {\it operational}. It turns out, however, that in case of conserving charge transfer through normal chaotic structures, it is sufficient to know a probability measure associated with a set ${\bm T}$ of non-zero transmission eigenvalues $\{T_j \in (0,1)\}$; these are the eigenvalues of the Wishart-type matrix ${\bm t}{\bm t}^\dagger$, where ${\bm t}$ is the transmission sub-block of the reflection-transmission decomposed scattering matrix
\begin{eqnarray}\label{s-block}
    \bm{{\mathcal S}} = \left(
                      \begin{array}{cc}
                 {\bm r}_{N_{\rm L}\times N_{\rm L}} & {\bm t}_{n_{\rm L}\times N_{\rm R}} \\
                 {\bm t^\prime}_{N_{\rm R}\times N_{\rm L}} & {\bm r^\prime}_{N_{\rm R}\times N_{\rm R}} \\
               \end{array}
                   \right).
\end{eqnarray}
Owing to this observation, the joint probability density function of all transmission eigenvalues emerges as the object of primary interest in the random matrix theories (RMT) of quantum transport which, besides, are particularly suited to the studies of {\it integrable aspects} of the universal quantum transport.

The present paper reviews recent advances in the {\it integrable theory} of quantum transport in chaotic structures, making particular emphasis on establishing exact relations between experimentally measurable fluctuations of transport observables (such as the Landauer conductance and the noise power) and Painlev\'e and Toda Lattice equations. It should be stressed that the above relations become manifest only at the level of a moment generating function; this explains, to some extent, why they have been overlooked in the studies of other groups that have merely focussed on a nonperturbative calculation of the moments or cumulants of transport observables [see, e.g., Savin and Sommers (2006), Savin, Sommers and Wieczorek (2008), Novaes (2008), Khoruzhenko, Savin and Sommers (2009), Mezzadri and Simm (2011)]. Our exposition is mainly based on the results obtained by Osipov and Kanzieper (2008, 2009) and Vidal and Kanzieper (2012) who have considered the simplest, $\beta=2$ symmetry class referring to chaotic cavities with broken time-reversal symmetry. For discussion of integrability for $\beta=1$ and $\beta=4$ Dyson's symmetry classes, the reader is referred to the recent paper by Mezzadri and Simm (2013).

In Sections \ref{SoLC} and \ref{itnf}, chaotic cavities probed via two {\it ideal leads} [$\Gamma_j=1$ in Eq. (\ref{gj})] are considered. In this case, stochastic ${\bm {\mathcal S}}$-matrix description becomes particularly simple: the vanishing average scattering matrix ${\bm {\mathcal S}}_0={\bm 0}$ gives rise to the flat measure $P({\bm {\mathcal S}})=1$ which implies that scattering matrices
${\bm {\mathcal S}}$ belong to one of the three Dyson circular ensembles (Bl\"umel and Smilansky 1990, Lewenkopf and Weidenm\"uller 1991,
Brouwer 1995, Mello and Baranger 1999). By combining this observation with the ideas of integrability [see, e.g., Morozov (1994), Adler and van Moerbeke (2001), Osipov and Kanzieper (2010)], we show there that the problem of universal quantum
transport in chaotic cavities with broken time-reversal symmetry
($\beta=2$) is completely integrable. Although this conclusion is very general (Osipov and Kanzieper 2008, 2009) and applies
to a variety of transport observables, the discussion is purposely restricted to two particular problems listed in order of increasing difficulty: (i) statistics of the Landauer
conductance (Section \ref{SoLC}), and (ii) statistics of current fluctuations quantified by the noise power (Section \ref{itnf}). This will help us to keep the presentation as transparent
as possible and also show the merits of the integrability based approach. Emergence of (fifth) {\it Painlev\'e transcendents} and {\it one-dimensional Toda Lattices} in a nonperturbative description of quantum transport is the main outcome of Sections \ref{SoLC} and \ref{itnf}.

Integrable theory of quantum transport (Vidal and Kanzieper 2012) that accounts for tunneling effects in the point contacts [{\it non-ideal leads}, $0 < \Gamma_j <1$ in Eq. (\ref{gj})] is formulated in Section \ref{Tunneling}. There, we show that non-ideality of the leads does not destroy integrability of the problem. In particular, we prove that conductance fluctuations are governed by a {\it two-dimensional Toda Lattice} equation.

\section{Integrable theory of conductance fluctuations in chaotic cavities with ideal leads}\label{SoLC}
\subsection{Moment generating function for Landauer conductance}

In what follows, we consider chaotic cavities with broken
time-reversal symmetry which are probed, via ballistic point
contacts [no tunneling, $\Gamma_j=1$ in Eq. (\ref{gj})], by two (left and right) leads further attached to outside reservoirs kept at temperature $\theta$; the leads support
$N_{\rm L}$ and $N_{\rm R}$ propagating modes, respectively. In this scattering geometry, the Landauer conductance at zero temperature is related the scattering matrix ${\bm {\mathcal S}}$ of the system via Landauer formula
\begin{eqnarray}
    G = {\rm tr\,} ({\mathcal C}_1 {\bm {\mathcal S}}{\mathcal C}_2 {\bm {\mathcal S}}^\dagger),
\end{eqnarray}
where ${\mathcal C}_{1,2}$ are projection
matrices of the form
\begin{eqnarray} \label{c1-c2}
{\mathcal C}_1= \left(
                  \begin{array}{cc}
                    \mathds{1}_{N_{\rm L}} & 0 \\
                    0 & 0_{N_{\rm R}} \\
                  \end{array}
                \right),\;\;\;
{\mathcal C}_2= \left(
                  \begin{array}{cc}
                    0_{N_{\rm L}} & 0 \\
                    0 & \mathds{1}_{N_{\rm R}} \\
                  \end{array}
                \right).
\end{eqnarray}

In order to describe fluctuations
of the conductance $G = {\rm tr\,} ({\mathcal C}_1 {\bm{\mathcal S}}
{\mathcal C}_2 {\bm {\mathcal S}}^\dagger)$ in an adequate way, one needs
to know its entire distribution function. To determine the latter,
we define the moment generating function (MGF)
\begin{eqnarray}
\label{c-iz}
    {\mathcal F}_n^{(\nu)}(z) = \left<
        \exp\left(
            - z G \right)
    \right>_{{\bm {\mathcal S}}\in {\rm CUE}(N_{\rm L}+N_{\rm R})}
\end{eqnarray}
labeled by the indices
\numparts
\begin{equation}\label{n-index}
n=\min(N_{\rm L},N_{\rm R}),
\end{equation}
and
\begin{equation} \label{nu-index}
\nu=|N_{\rm L}-N_{\rm R}|,
\end{equation}
\endnumparts
the latter being the asymmetry parameter (we shall occasionally omit the superscript $(\nu)$ wherever this does not cause a notational confusion). The logarithm, $\log {\mathcal F}_n^{(\nu)}(z)$, Taylor-expanded in a vicinity of $z=0$, supplies the cumulants of Landauer conductance.

While the averaging in Eq.~(\ref{c-iz}), running over scattering matrices\linebreak ${\bm {\mathcal S}}\in {\rm CUE}(N_{\rm L}+N_{\rm R})$ drawn
from the Dyson circular unitary ensemble, can explicitly be performed
with the help of the Harish-Chandra-Itzykson-Zuber formula (Harish-Chandra 1957; Itzykson and Zuber 1980), a high
spectral degeneracy of the projection matrices ${\mathcal C}_1$ and
${\mathcal C}_2$ makes this calculation quite tedious. To avoid
unnecessary technical complications, it is beneficial to decompose the scattering matrix into reflection-transmission submatrices [see Eq.~(\ref{s-block})]
to realise that the Landauer conductance $G$ is solely determined by (transmission) eigenvalues $\{ T_j \}$ of ${\bm t}{\bm t}^\dagger$ (Landauer 1957, Fisher and Lee 1981, B\"uttiker 1990),
\begin{eqnarray}
\label{T-cond}
    G({\bm T}) = {\rm tr\,} ({\bm t}{\bm t}^\dagger) = \sum_{j=1}^n T_j.
\end{eqnarray}
The uniformity of the scattering ${\mathcal S}$-matrix distribution
gives rise to a nontrivial joint probability density function of
transmission eigenvalues in the form (Baranger and Mello 1994; Jalabert, Pichard and Beenakker 1994; Forrester 2006)
\begin{eqnarray}
\label{PnT}
    P_n^{(\nu)}({\bm T}) = c_{n,\nu}^{-1} \,\Delta_n^2({\bm T}) \prod_{j=1}^n
    T_j^\nu.
\end{eqnarray}
Here, $\Delta_n({\bm T})=\prod_{j<k} (T_k-T_j)$ is the Vandermonde
determinant and $c_{n,\nu}$ is the normalisation constant (Mehta 2004)
\begin{eqnarray}
\label{nc}
    c_{n,\nu} = \prod_{j=0}^{n-1} \frac{\Gamma(j+2)\, \Gamma(j+\nu+1) \,\Gamma(j+1)}
    {\Gamma(j+\nu+n+1)}.
\end{eqnarray}
Let us stress that the description based on Eqs.~(\ref{T-cond}) and (\ref{PnT}) is completely
equivalent to the original ${\bm {\mathcal S}}\in {\rm CUE}(N_{\rm L}+N_{\rm R})$ model microscopically justified (Lewenkopf and Weidenm\"uller 1991; Brouwer 1995) in the universal transport regime we are confined to. Equation (\ref{PnT}) is one of the cornerstones of the RMT approach to quantum transport.

\subsection{Non-perturbative calculation of the moment generating function (easy way)}\label{Easy-Way}
Appearance of the fifth Painlev\'e transcendent in the Landauer conductance moment generating function (MGF), announced in the abstract to this contribution, can be appreciated in an ``easy way'' after one realises that the matrix/eigenvalue integral \footnote{Essentially the same eigenvalue integral appears in the replica approach to {\it statical} correlations in the Calogero-Sutherland model with the interaction parameter $\lambda=1$, see, e.g., Gangardt and Kamenev (2001) and Kanzieper (2002).}
\begin{eqnarray}
\label{cond-eig-nod}
    {\mathcal F}_n^{(\nu)}(z) = c_{n,\nu}^{-1} \int_{(0,1)^n} \prod_{j=1}^n dT_j\, T_j^\nu \exp(-zT_j)\cdot \Delta_n^2({\bm T})
\end{eqnarray}
can be described in terms of completely integrable Toda Lattice model (Toda 1989; Teschl 2000). The Darboux theorem (Darboux 1889; Vein and Dale 1999) followed by the Toda-to-Painlev\'e reduction (Okamoto 1987; Forrester and Witte 2002) are the key ingredients of the calculation to be presented below.

\subsubsection{MGF and the Toda Lattice equation}\label{TL-Hankel}
\noindent\newline\newline
A close inspection of the integral Eq.~(\ref{cond-eig-nod}) reveals that it admits the Hankel determinant representation (Osipov and Kanzieper 2008)
\begin{eqnarray}
\label{hd}
    {\mathcal F}_n^{(\nu)}(z) =  \frac{n!}{c_{n,\nu}} \, {\rm det}}\left[
        (-\partial_z)^{j+k}\, {\mathcal F}_1^{(\nu)}(z)
    \right]_{(j,k)\in (0,n-1)
\end{eqnarray}
with
\begin{equation}
\label{f1}
 {\mathcal F}_1^{(\nu)}(z) =\int_0^1 dT\, T^{\nu} \,\exp(-zT) = \frac{\Gamma(\nu+1)}{z^{\nu+1}}
    \left(
            1 - e^{-z} \sum_{\ell=0}^\nu \frac{z^\ell}{\ell!}
    \right).
\end{equation}
In deriving Eqs.~(\ref{hd}) and (\ref{f1}) we have used the
Andr\'eief--de Bruijn integration formula (Andr\'eief 1883, de Bruijn 1955).

Equation (\ref{hd}), supplemented by the ``initial condition''
${\mathcal F}_0^{(\nu)}(z)=1$, has far-reaching consequences. Indeed, by
virtue of the Darboux theorem (Darboux 1889; Vein and Dale 1999), the infinite sequence
of the moment generating functions $\{{\mathcal F}_1^{(\nu)},{\mathcal
F}_2^{(\nu)},\cdots\}$ obeys the {\it one-dimensional Toda Lattice} equation $(n\ge 1)$
\begin{equation}
\label{TL}
\hspace{-1cm}
    {\mathcal F}_n^{(\nu)}(z)\,\frac{\partial^2 {\mathcal F}_n^{(\nu)}(z)}{\partial z^2} - \left(
    \frac{\partial {\mathcal F}_n^{(\nu)}(z)}{\partial z}\right)^2 = {\rm var}_{n,\nu}(G)\,
    {\mathcal F}_{n-1}^{(\nu)}(z)\, {\mathcal F}_{n+1}^{(\nu)}(z),
\end{equation}
where ${\rm var}_{n,\nu}(G) = n(n+1)^{-1} (c_{n-1,\nu} c_{n+1,\nu}/ c_{n,\nu}^2)$ is, by definition,
the conductance variance
\begin{equation}
\label{varG}
    {\rm var}_{n,\nu}(G) =  \frac{n^2 (n+\nu)^2}{(2n+\nu)^2
    [(2n+\nu)^2-1]}.
\end{equation}
(To appreciate this point, project the Toda Lattice equation onto $z=0$.) Importantly, emergence of the Toda Lattice equation for the MGF can be traced back to the broken time-reversal symmetry in the underlying scattering system. For chaotic cavities with conserved time-reversal symmetry ($\beta=1$) and in presence of spin-orbit interactions ($\beta=4$), the statistics of conductance fluctuations is captured by the Pfaff-KP Lattice equation (Mezzadri and Simm 2013).
\noindent \newline\newline
Since ${\mathcal F}_n^{(\nu)}(z)$ is the Laplace transform of conductance
probability density\linebreak $f_n^{(\nu)}(g)=\langle \delta (g - G) \rangle$, the
Toda lattice equation provides an exact recursive solution to the
problem of conductance distribution in chaotic cavities with an
arbitrary number of channels in the leads. Indeed, because of a specific form of ${\mathcal F}_1^{(\nu)}(z)$, a calculation of the inverse Laplace transform of
    ${\mathcal F}_n^{(\nu)}(z)$ is operationally straightforward. The resulting
    probability density function $f_n^{(\nu)}(g)$ can be shown to be a nonanalytic function, admitting the representation:
\begin{eqnarray}
    f_n^{(\nu)}(g) = \sum_{k=0}^n  {\rm sgn} (g-k) \, \pi_{k}^{(n,\nu)}(g-k),\qquad g\in(0,n).
\end{eqnarray}
Here, $\{\pi_{0}^{(n,\nu)}(x), \pi_{1}^{(n,\nu)}(x),\dots, \pi_{n}^{(n,\nu)}(x)\}$ is a set of hierarchically structured polynomials of degree ${\rm deg\,}\pi_{k}^{(n,\nu)}(x)=n(n+\nu)-1$, satisfying a number of remarkable properties that originate from the structures hiding behind the Toda Lattice Eq.~(\ref{TL}). For $\nu=0$, explicit formulae for  $\pi_{k}^{(n,0)}(x)$ ($1\le n \le 4$) are given in Table \ref{pol-table}. Finding explicit expressions for these polynomials for generic $n$ and $\nu$ is a nontrivial problem.

\begin{table}
\caption{\label{pol-table} Polynomials $\pi_{k}^{(n,0)}(x)$ obtained for $1\le n \le 4$ by iterating the Toda Lattice equation (\ref{TL}) and further applying
the inverse Laplace transform. Due to the relation
$$
\pi_{n-k}^{(n,0)}(x)=-\pi_k^{(n,0)}(-x)
$$
holding in the symmetric case ($\nu=0$) presented in the Table, it is sufficient to calculate $\pi_{k}^{(n,0)}(x)$ for $k \le [n/2]$.}

\begin{indented}
\lineup
\item[]
\begin{tabular}{@{}*{2}{l}} \br
$\0\0 n$& $\pi_k^{(n,0)}(x)$ \cr \mr

$\0\0 1$& $\pi_0^{(1,0)}(x) = \displaystyle{\frac{1}{2}}$ \vspace{0.1cm} \cr \mr

$\0\0 2$& $\pi_0^{(2,0)}(x) = \phantom{-2}x^3$\vspace{0.1cm} \cr

$\0\0 {}$& $\pi_1^{(2,0)}(x) = -2 x\, (x^2+3)$\vspace{0.1cm} \cr \mr

$\0\0 3$& $\pi_0^{(3,0)}(x) = \phantom{-}\displaystyle{\frac{3}{28} x^8}$ \vspace{0.1cm}\cr

$\0\0 {}$& $\pi_1^{(3,0)}(x) = \displaystyle{-\frac{9}{28} \, x^4 \left( x^4 + 56 x^2 -112 x + 140  \right)}$ \vspace{0.1cm}\cr \mr

$\0\0 4$& $\pi_0^{(4,0)}(x) = \phantom{-}\displaystyle{\frac{1}{3003} x^{15}}$ \vspace{0.1cm}\cr

$\0\0 {}$& $\pi_1^{(4,0)}(x) = \displaystyle{-\frac{4}{3003} x^{9}} \left( x^6 + 315 x^4 - 2730 x^3 +15015 x^2 -30030 x + 25025 \right)$ \vspace{0.1cm}\cr

$\0\0 {}$& $\pi_2^{(4,0)}(x) = \displaystyle{-\frac{2}{3003} x^{7}} \left(3 x^8 + 1260 x^6 + 10920 x^4 + 400400 x^2 +900900 \right)$ \vspace{0.1cm}\cr

\br
\end{tabular}
\end{indented}
\end{table}

\subsubsection{MGF and the Painlev\'e V equation}\noindent\newline\newline
{\it (i) Toda-to-Painlev\'e reduction.}---While important from conceptual point of view and also operationally useful in generating explicit formulae for the {\it distribution} of Landauer conductance for finite $n$, the Toda Lattice equation is of little help in studying the conductance cumulants. Fortunately, the Toda Lattice representation of the MGF can readily be converted into a Painlev\'e representation since, miraculously, the same Toda Lattice equation governs the behaviour of so-called $\tau$
functions arising in the Hamiltonian formulation of the six Painlev\'e equations (Clarkson 2003; Noumi 2004), which are yet another fundamental object in the theory of nonlinear
integrable lattices. The aforementioned Painlev\'e reduction (Okamoto 1987; Forrester and Witte 2002) of the Toda Lattice equation (\ref{TL}) materialises in the exact representation
\begin{eqnarray}
\label{FnP-22}
    {\mathcal F}_n^{(\nu)}(z) = \exp\left(
        \int_0^z dt \frac{\sigma_{\rm V}(t; n,\nu) - n(n+\nu)}{t}
    \right).
\end{eqnarray}
Here, $\sigma_{\rm V}(t;n,\nu)=\sigma_{n,\nu}(t)$ satisfies the Jimbo-Miwa-Okamoto form of
the {\it Painlev\'e V} equation
\begin{eqnarray}
\label{pv-22} \hspace{-1cm}
    (t \sigma_{n,\nu}^{\prime\prime})^2 - [\sigma_{n,\nu} - t \sigma_{n,\nu}^\prime
    &+& 2 (\sigma_{n,\nu}^\prime)^2 + (2n+\nu)\sigma_{n,\nu}^\prime]^2 \nonumber\\
     &+& 4(\sigma_{n,\nu}^\prime)^2 (\sigma_{n,\nu}^\prime + n) (\sigma_{n,\nu}^\prime+n+\nu)=0
\end{eqnarray}
subject to the boundary condition
\begin{eqnarray}\label{pv-bc}
\sigma_{n,\nu}(t\rightarrow
0)\simeq n(n+\nu) - \frac{n(n+\nu)}{2n+\nu}t + {\mathcal O}(t^2).
\end{eqnarray}

\noindent\newline\newline
{\it (ii) MGF as a gap formation probability.}---There exists yet another, ``easy'' way to derive the Painlev\'e V representation of the MGF. Spotting that the moment generating function ${\mathcal F}_n^{(\nu)}(z)$ is essentially a gap formation probability $E_{n,\nu}^{{\rm LUE}}(0; (z,\infty))$ within the interval $(z,+\infty)$ in the spectrum of
an auxiliary $n \times n$ Laguerre unitary ensemble (Mehta 2004),
\begin{eqnarray}
    {\mathcal F}_n^{(\nu)}(z) =z^{-n(n+\nu)}\, E_{n,\nu}^{{\rm LUE}}(0; (z,\infty)),
\end{eqnarray}
where
\begin{eqnarray}
    E_{n,\nu}^{{\rm LUE}}(0; (z,\infty)) = c_{n,\nu}^{-1}\int_{(0,z)^n}
    \prod_{j=1}^n d\lambda_j \, \lambda_j^\nu \, e^{-\lambda_j}
    \cdot \Delta_n^2({\bm \lambda}),
\end{eqnarray}
one immediately reproduces (Tracy and Widom 1994) Eqs.~(\ref{FnP-22}) and (\ref{pv-22}).
\noindent\newline\newline
The Painlev\'e~V representation of the MGF (Osipov and Kanzieper 2008) opens a way for an elegant nonperturbative calculation of the Landauer conductance
cumulants (see Section \ref{cum-section}). As a matter of fact, Eq.~(\ref{FnP-22}) incorporates all available nonperturbative results for cumulants of the Landauer conductance at $\beta=2$ (see, e.g., Savin and Sommers 2006; Savin, Sommers and Wieczorek 2008; Novaes 2008; Khoruzhenko, Savin and Sommers (2009)).

\subsection{The $\tau$ function theory of the moment generating function}
The treatment presented in Section \ref{Easy-Way} was largely based on a wealth of ``ready-for-use'' results (Andr\'eief-de Bruijn formula, Darboux theorem, a connection between the Toda Lattice and Painlev\'e transcendents, and a well-studied problem of calculating gap formation probabilities) which surprisingly well fitted our goal of a nonperturbative
evaluation of the particular moment generating function Eq.~(\ref{cond-eig-nod}). Since existence of such an ``easy way'' is clearly the exception rather than the rule (see, e.g., Section \ref{itnf} where statistics of thermal-to-shot-noise crossover in chaotic cavities is studied), a regular yet flexible formalism is needed for a nonperturbative description of a wide class of generating
functions arising in the context of universal quantum transport in chaotic  structures. To outline such a formalism, we first revisit the problem of a nonperturbative evaluation of the Landauer conductance MGF in order to reproduce the main results of the previous Section in a coherent manner.

\subsubsection{The idea}\noindent\newline\newline
The ``deform-and-study'' approach (Morozov 1994; Adler, Shiota and van Moerbeke 1995; Adler and van Moerbeke 1995, 2001; Osipov and Kanzieper 2010)  borrowed from the theory of
integrable systems is central to the nonperturbative calculation of
${\mathcal F}_n^{(\nu)}(z)$. In the present context, the main idea of the
method is ``embedding'' ${\mathcal F}_n^{(\nu)}(z)$ into a more general
theory of the $\tau$ function
\begin{equation}
    \label{tau-def-00}
    \tau_n^{(\nu)}({\bm t};z) = \frac{1}{n!}\int_{(0,1)^n} \prod_{j=1}^n dT_j\,T_j^\nu \,\Gamma_{z}(T_j)
    \,e^{V({\bm t};T_j)} \Delta_n^2({\bm T})
\end{equation}
which possesses the infinite-dimensional parameter space ${\bm
t}=(t_1,t_2,\dots)$ arising as the result of the ${\bm t}$
deformation \footnote{Here, the notation ${\bm t}$ should not be confused with the one previously used for the transmission sub-block of the scattering matrix ${\bm {\mathcal S}}$, Eqs. (\ref{s-block}) and (\ref{T-cond}). }
\begin{eqnarray}
V({\bm t};T) = \sum_{k=1}^\infty t_k T^k
\end{eqnarray}
of the MGF of a generic transport observable [see, e.g., Eq.~(\ref{cond-eig-nod})]. In the context of the Landauer conductance fluctuations, the function $\Gamma_{z}(T)$ should be set to $\Gamma_{z}(T)=\exp(-z T)$.

Studying an evolution of the $\tau$ function in the extended $(n,
{\bm t},z)$ space allows us to identify various nonlinear
differential hierarchical relations. A projection of these relations
onto the hyperplane ${\bm t}={\bm 0}$,
\begin{eqnarray}
\label{proj-00}
    {\mathcal F}_n^{(\nu)}(z) = \frac{n!}{c_{n,\nu}}\,
    \tau_n^{(\nu)}({\bm t};z)\Big|_{{\bm t}={\bm 0}},
\end{eqnarray}
generates, among others, a closed nonlinear differential
equation for the moment generating function ${\mathcal F}_n^{(\nu)}(z)$.

The two key ingredients of the exact theory of $\tau$ functions are
(i) the bilinear identity (Date et al 1983) and (ii) the
(linear) Virasoro constraints (Mironov and Morozov 1990).

\subsubsection{Bilinear identity and integrable hierarchies}\noindent\newline\newline
The bilinear identity encodes an infinite set of hierarchically
structured nonlinear differential equations in the variables ${\bm
t}=(t_1, t_2,\dots)$. For the model introduced in
Eq.~(\ref{tau-def-00}), the bilinear identity reads (Adler, Shiota and van Moerbeke 1995; Osipov and Kanzieper 2010):
\begin{eqnarray} \label{bi-id}
    \oint_{{\cal C}_\infty} d^2\xi \,e^{a\, v(\bm{t-t^\prime};\xi)}
    \tau_{n}^{(\nu)}(\bm{t}-[\bm{\xi}^{-1}])\,
    \frac{\tau_{m+1}^{(\nu)}(\bm{t^\prime}+[\bm{\xi}^{-1}])}{\xi^{m+1-n}}\qquad
    \nonumber\\
    \qquad =\oint_{{\cal C}_\infty} d^2\xi \,e^{(a-1)\, v(\bm{t-t^\prime};\xi)}
    \tau_m^{(\nu)}(\bm{t^\prime}-[\bm{\xi}^{-1}]) \frac{\tau_{n+1}^{(\nu)}
    (\bm{t}+[\bm{\xi}^{-1}])}{\xi^{n+1-m}}.
\end{eqnarray}
Here, $\xi \in {\mathbb C}$ and $a\in {\mathbb R}$ is a free parameter; the integration
contour ${\cal C}_\infty$ encompasses the point $\xi=\infty$; the
notation ${\bm t} \pm [{\bm \xi}^{-1}]$ stands for the infinite set of
parameters $\{t_j\pm \xi^{-j}/j\}$; for brevity, both the Laplace parameter $z$ and the asymmetry index $\nu$ were dropped from the arguments of
$\tau$ functions.

Being expanded in terms of $\bm{ t^\prime}-{\bm t}$ and $a$,
Eq.~(\ref{bi-id}) generates a zoo of integrable hierarchies satisfied by the $\tau$ function Eq.~(\ref{tau-def-00}) in the $(n,{\bm t})$-space. The Toda Lattice (TL) and Kadomtsev-Petviashvili (KP) hierarchies are central to our approach. The first non-trivial members of the TL and KP hierarchies read~\footnote[2]{Notice that due to the identity
$$
\frac{\partial}{\partial t_1} \tau_n^{(\nu)}({\bm t};z) = - \frac{\partial}{\partial z} \tau_n^{(\nu)}({\bm t};z),
$$
Eq.~(\ref{ftl}) projected onto ${\bm t}={\bm 0}$ readily yields the Toda Lattice equation Eq.~(\ref{TL}) for the MGF ${\mathcal F}_n^{(\nu)}(z)$ claimed in Section 2.2.1 with the help of Darboux theorem.
}
\begin{eqnarray}\label{ftl}
    \tau_n^{(\nu)}({\bm t})\, \frac{\partial^2 \tau_n^{(\nu)}({\bm t})}{\partial t_1^2} - \left(
        \frac{\partial\tau_n^{(\nu)}({\bm t})}{\partial t_1}
    \right)^2 = \tau_{n-1}^{(\nu)}({\bm t})\,  \tau_{n+1}^{(\nu)}({\bm t})
\end{eqnarray}
and
\begin{eqnarray}\fl
    \label{fkp}
        \left(
    \frac{\partial^4}{\partial t_1^4} + 3\,\frac{\partial^2}{\partial t_2^2} -
        4\, \frac{\partial^2}{\partial t_1 \partial t_3}
    \right)\, \log \tau_n^{(\nu)}({\bm t})
    + \,6\, \left(
        \frac{\partial^2}{\partial t_1^2}\, \log \tau_n^{(\nu)}({\bm t})
    \right)^2 = 0,
\end{eqnarray}
respectively (Adler, Shiota and van Moerbeke 1995; Adler and van Moerbeke 1995). For higher-order equations of the TL and KP hierarchies, as well as a complete list of integrable hierarchies encoded in the bilinear identity Eq.~(\ref{bi-id}), the reader is referred to Osipov and Kanzieper (2010).

\subsubsection{Virasoro constraints}\noindent\newline\newline
The projection formula Eq.~(\ref{proj-00}) makes it tempting to assume that nonlinear integrable hierarchies
satisfied by $\tau$ functions in the $(n, {\bm t})$-space should
induce similar, hierarchically structured, nonlinear differential
equations for the moment generating function ${\mathcal F}_n^{(\nu)}(z)$. To identify them, one has to seek an
additional block of the theory that would make a link between the partial
$\{t_k\}$ derivatives of $\tau$ functions taken
at ${\bm t}={\bm 0}$ and the derivatives of ${\mathcal F}_n^{(\nu)}(z)$ with respect to the Laplace parameter $z$. The study by Adler,
Shiota and van Moerbeke (1995) suggests that the missing block is
the {\it Virasoro constraints} which reflect
invariance of the $\tau$ function [Eq.~(\ref{tau-def-00})] under a
change of the integration variables.

In the present context, it is useful to demand the invariance under
the set of transformations
\begin{eqnarray}
\label{v-tr}
T_j \rightarrow \tilde{T}_j +
    \epsilon \,\tilde{T}_j^{q+1} (\tilde{T}_j-1),\;\;\; q\ge 0
\end{eqnarray}
that leave the integration domain intact. Employing a by now standard procedure (Mironov and Morozov 1990; Adler and van Moerbeke 1995; Osipov and Kanzieper 2010),
one readily checks that the transformation (\ref{v-tr}) induces the Virasoro
constraints in the form
\begin{equation}
\label{vc-1}
    \big[ \hat{L}_{q+1}({\bm t}) - \hat{L}_{q}({\bm t})\big] \tau_n^{(\nu)}({\bm t};z)
     = 0,\;\;\;  q\ge 0,
\end{equation}
where a set of differential operators
\begin{eqnarray}
\label{vc-2-00}
    \hat{L}_q({\bm t})
     =
     \hat{\mathcal L}_{q}({\bm t}) - z\, \frac{\partial}{\partial t_{q+1}}
        + \nu \frac{\partial}{\partial t_q}
\end{eqnarray}
involves the Virasoro operators
\begin{eqnarray}
    \label{vc-3}
    \hat{{\cal L}}_q({\bm t}) = \sum_{j=1}^\infty jt_j \,\frac{\partial}{\partial t_{q+j}}
    +
    \sum_{j=0}^q \frac{\partial^2}{\partial t_j \partial t_{q-j}},
\end{eqnarray}
satisfying the Virasoro algebra
\begin{eqnarray}\label{va}
    [\hat{{\cal L}}_p,\hat{{\cal L}}_q] = (p-q)\hat{{\cal
    L}}_{p+q}, \;\;\;\; p,q\ge -1.
\end{eqnarray}
Equations~(\ref{vc-2-00}) and (\ref{vc-3}) assume the convention
$\partial/\partial t_0 \equiv n$.

\subsubsection{The Toda Lattice equation for MGF}\noindent\newline\newline
The Toda Lattice equation for the MGF ${\mathcal F}_n^{(\nu)}(z)$ follows from the projection formula Eq.~(\ref{proj-00}), the TL equation (\ref{ftl}) written in the $(n,{\bm t})$-space, and the relation
\begin{eqnarray}\label{t1z}
\frac{\partial}{\partial t_1} \tau_n^{(\nu)}({\bm t};z) = -
\frac{\partial}{\partial z} \tau_n^{(\nu)}({\bm t};z).
\end{eqnarray}
Straightforward manipulations bring out Eqs.~(\ref{TL}) and (\ref{varG}), obtained in Section \ref{TL-Hankel} by virtue of the Darboux theorem applied to a Hankel determinant form of the MGF.

\subsubsection{The Painlev\'e V equation for MGF}\noindent\newline\newline
To derive a differential equation for the MGF ${\mathcal
F}_n(z)$, we combine the projection formula Eq.~(\ref{proj-00}) with the first KP equation (\ref{fkp}). Having in mind the identity Eq.~(\ref{t1z}) as well as the two lowest Virasoro constraints labeled by
$q=0$,
\begin{eqnarray} \fl \label{q0-00}
    \left[
    \sum_{j=1}^\infty jt_j \left(
    \frac{\partial}{\partial t_{j+1}} - \frac{\partial}{\partial t_{j}}\right)
    - z \frac{\partial}{\partial t_2} +(2n+\nu+z) \frac{\partial}{\partial t_1}
    \right] \log \tau_n^{(\nu)}({\bm t}; z)=n(n+\nu),
\end{eqnarray}
and $q=1$
\begin{eqnarray} \fl \label{q1-00}
    \Bigg[
    \sum_{j=1}^\infty jt_j \left(
    \frac{\partial}{\partial t_{j+2}} - \frac{\partial}{\partial t_{j+1}}\right)
    -z \frac{\partial}{\partial t_3} + (2n+\nu+z) \frac{\partial}{\partial t_2}\nonumber\\
     - (2n+\nu) \frac{\partial}{\partial t_1}
    + \frac{\partial^2}{\partial t_1^2}\Bigg] \log \tau_n^{(\nu)}({\bm t}; z) + \left(
    \frac{\partial}{\partial t_1} \log \tau_n^{(\nu)}({\bm t},z)
    \right)^2=0,
\end{eqnarray}
we reveal, after lengthy but straightforward manipulations, that the function
\begin{eqnarray}
\label{sF-cc}
    \sigma_{n,\nu}(z) = n(n+\nu) + z \frac{\partial}{\partial z}\log {\mathcal F}_n^{(\nu)}(z)
\end{eqnarray}
satisfies a nonlinear differential equation
\begin{eqnarray}\fl
\label{OK-2008}
    z^2 \sigma_{n,\nu}^{\prime\prime\prime} +
    z \,\sigma_{n,\nu}^{\prime\prime} +
    6 z \left(\sigma_{n,\nu}^\prime
    \right)^2
    - 4 \sigma_{n,\nu} \sigma_{n,\nu}^\prime \nonumber \\
    -
        \left[
            z^2 -2(2n +\nu)\,z+\nu^2        \right]\,
  \sigma_{n,\nu}^\prime -
    (2n+\nu -z) \sigma_{n,\nu} = 0
\end{eqnarray}
belonging to the Chazy I class (Chazy 1911; Cosgrove and
Scoufis 1993).  This can be recognised as the Chazy form of the fifth Painlev\'e transcendent [Eq.~(\ref{pv-22})]
written in the Jimbo-Miwa-Okamoto form (Jimbo et al 1980; Okamoto 1987). Equation (\ref{FnP-22}) then readily follows. For transition from Eq.~(\ref{OK-2008}) to Eq.~(\ref{pv-22}), see Appendix E in the paper by Osipov and Kanzieper (2010).

\subsection{Cumulants of the Landauer conductance}\label{cum-section}
The Painlev\'e V representation of the MGF, Eq.~(\ref{FnP-22}), opens a way for a nonperturbative calculation of the Landauer conductance cumulants.

\subsubsection{Cumulants from Taylor expanded Painlev\'e V}\noindent\newline\newline
Perhaps, the most startling consequence of the above nonperturbative calculation is the claim (Osipov and Kanzieper 2008) that
the cumulants of the Landauer conductance of a cavity probed via two ideal leads are furnished by the coefficients of a Taylor expanded Painlev\'e V function. Indeed, since the cumulants $\{\kappa_\ell^{(n,\nu)} (G)\}$ are supplied by the moment generating function
\begin{eqnarray}
    \log {\mathcal F}_n^{(\nu)}(z) = \sum_{\ell=1}^\infty \frac{(-1)^\ell}{\ell!} \,
    \kappa_\ell^{(n,\nu)}(G)\, z^\ell,
\end{eqnarray}
one derives, in view of Eq.~(\ref{sF-cc}), the remarkable identity:
\begin{eqnarray}\label{main-cond-cum}
    \sigma_{n,\nu}(z) = n(n+\nu) + \sum_{\ell=1}^\infty \frac{(-1)^\ell}{\Gamma(\ell)}\, \kappa_\ell^{(n,\nu)}(G)\, z^\ell.
\end{eqnarray}
Whenever the cumulant dependence on $n$ and $\nu$ is clear from the context, the superscript ${(n,\nu)}$ will be omitted.

\subsubsection{Exact recurrence solution}\noindent\newline\newline
 Alternatively, the cumulants of any finite order can be generated by a recurrence relation. The latter follows upon substitution of the expansion Eq.~(\ref{main-cond-cum}) into Chazy's form of the fifth Painlev\'e transcendent [Eq.~(\ref{OK-2008})]. The resulting recurrence relation reads~\footnote[1]{Notice that Eq.~(15) in the paper by Osipov
and Kanzieper (2008) contains typos. The correct formula is given by
Eq.~(\ref{cumeq}).}:
\begin{eqnarray}
\label{cumeq}
\fl [(2n+\nu)^2-\ell^2]\,(\ell+1) \kappa_{\ell+1}^{(n,\nu)}(G)
        + (2n+\nu) (2\ell-1)\,\ell \,\kappa_{\ell}^{(n,\nu)}(G) \nonumber\\
        \fl \qquad + \ell(\ell-1)(\ell-2)\,
     \kappa_{\ell-1}^{(n,\nu)}(G)
     - 2 \sum_{j=0}^{\ell-1} (3j+1) (j-\ell)^2 \left({\ell}\atop{j}\right)
     \kappa_{j+1}^{(n,\nu)}(G)\, \kappa_{\ell-j}^{(n,\nu)}(G) = 0,\nonumber\\
     {}
\end{eqnarray}
where $\ell \ge 2$, and
\begin{eqnarray}
    \left({\ell}\atop{j}\right) = \frac{\Gamma(\ell+1)}{\Gamma(j+1) \Gamma(\ell-j+1)}.
\end{eqnarray}
Taken together with the
initial conditions provided by the mean conductance $\kappa_1(G)$ and its variance $\kappa_2(G)$ (Baranger and Mello 1994, Jalabert, Pichard and Beenakker 1994),
\begin{eqnarray} \label{ic-cond}
\kappa_1(G) = \frac{n (n+\nu)}{2n + \nu},\quad
\kappa_2(G) = \frac{\kappa_1^2(G)}{(2n+\nu)^2 -1},
\end{eqnarray}
this recurrence efficiently generates conductance cumulants of any given order. Since the resulting expressions are very lengthy, we only quote the third cumulant related to the conductance skewness:
\begin{eqnarray}
    \kappa_{3}(G) = -\frac{\nu^2}{(2n+\nu)}\, \kappa_{1}^2(G).
\end{eqnarray}
This reproduces the result first obtained by Savin, Sommers and Wieczorek (2008) who used a Selberg integral technique; for an alternative calculation
based on the theory of symmetric functions, see Novaes (2008) and Khoruzhenko, Savin and Sommers (2009). Higher order cumulants can readily be deduced from Eq. (\ref{cumeq}) as well.

\subsubsection{Asymptotic analysis of the cumulants of finite order}\label{cum-fo}\noindent\newline\newline
Below, the Painlev\'e solution for the MGF will be studied in the limit of a large number of propagating modes in both left and right leads. The asymptotic analysis of conductance cumulants to be performed in this Section, refers to chaotic cavities in which the asymmetry parameter $\nu = |N_{\rm L} - N_{\rm R}|$ is kept {\it fixed} whilst $n={\rm min}(N_{\rm L},N_{\rm R})$ is assumed to be large, $n \gg 1$. The forthcoming calculation turns out to be particularly simple if one deals with the recurrence equation written in terms of
\begin{eqnarray}
    \chi_\ell = \frac{(-1)^\ell}{\Gamma(\ell)}\, \kappa_\ell(G).
\end{eqnarray}
From Eq.~(\ref{cumeq}), one derives:
\begin{eqnarray}\fl
    \label{rec-sigma}
    \phantom{00000000}
    (\ell+1)\big[\ell^2 - (2n+\nu)^2\big]\chi_{\ell+1} &+& (2n+\nu) (2\ell-1)\chi_\ell -(\ell-2)\chi_{\ell-1}\nonumber\\
    &=&
    - 2 \sum_{j=0}^{\ell-1}(\ell -j)(3j+1) \chi_{j+1} \chi_{\ell-j}.
\end{eqnarray}
The initial conditions for this equation, that holds down to $\ell=0\;$\footnote{Both the series $\sum_0^{-1}$ and the coefficient $\chi_{-1}$ should be set to zero.}, read:
\begin{eqnarray}
    \chi_0 = n(n+\nu),\qquad \chi_1 = -\frac{n(n+\nu)}{2n+\nu}=-\frac{n}{2} - \frac{\nu}{4} +{\mathcal O}(n^{-1}).
\end{eqnarray}
To develop a regular $1/n$ expansion for the set $\{\chi_\ell\}$, we separate therein the terms that do not vanish as $n\rightarrow \infty$,
\begin{eqnarray}
\label{sigma-ser}
    \chi_\ell = n(n+\nu) \delta_{\ell,0} - \frac{2n+\nu}{4} \delta_{\ell,1} +\frac{1}{16} \delta_{\ell,2} + \delta\chi_\ell \,\cdot {\mathds 1}_{\ell\ge 1},
\end{eqnarray}
and put forward the ansatz:
\begin{eqnarray}
\label{sigma-ans}
    \delta\chi_\ell =  \frac{a_\ell(\nu)}{(4n)^{\ell}} + \frac{b_\ell(\nu)}{(4n)^{\ell+1}}  + \frac{c_\ell(\nu)}{(4n)^{\ell+2}}
    + {\cal O}(n^{-\ell-3}).
\end{eqnarray}
In Eq. (\ref{sigma-ser}), the fraction $1/16$ originates from a non-vanishing (as $n\rightarrow \infty$) term of the $1/n$ expansion for
\begin{eqnarray}
    \chi_2 = \frac{n^2 (n+\nu)^2}{(2n+\nu)^2 [(2n+\nu)^2-1]} = \frac{1}{16} +{\mathcal O}(n^{-2}).
\end{eqnarray}
Substituting Eqs. (\ref{sigma-ser}) and (\ref{sigma-ans}) into Eq.~(\ref{rec-sigma}), the following set of recurrence equations can readily be derived from the resulting $1/n$ expansion:
\noindent\newline\newline
{\it (i)} The $\{a_\ell\}$--sequence:
\begin{eqnarray}
    a_{\ell+1}(\nu) = a_{\ell-1}(\nu),
\end{eqnarray}
where
\begin{eqnarray}
    a_1(\nu)=\frac{\nu^2}{2},\qquad a_2(\nu)=\frac{1-2\nu^2}{4}.
\end{eqnarray}
Hence, we deduce the solution
\begin{eqnarray}
\label{a-un-fa}
    a_\ell(\nu) = \frac{1}{8} \left[
        1 + (-1)^\ell (1-4\nu^2)\right].
\end{eqnarray}
\noindent\newline\newline
{\it (ii)} The $\{b_\ell\}$--sequence:
\begin{eqnarray}
    b_{\ell+1}(\nu) = b_{\ell-1}(\nu)-4\nu \, a_{\ell+1}(\nu).
\end{eqnarray}
This brings, after some effort,
\begin{eqnarray}
\label{a-un-fb}
    b_\ell(\nu) = -\ell\,\frac{\nu}{4} \left[
        1 + (-1)^\ell (1-4\nu^2) \right].
\end{eqnarray}
Equivalently,
\begin{eqnarray}
\label{a-un-fb-eq}
    b_\ell(\nu) = -2\nu\,\ell a_\ell(\nu).\\
    {}\nonumber
\end{eqnarray}
Combining the above results together, we derive the first few terms in the $1/n$ expansion of $\ell$-th conductance cumulant:
\begin{eqnarray}\fl\label{g-exp-1n}
    \kappa_\ell(G) =\frac{2n+\nu}{4}\delta_{\ell,1}+\frac{1}{16}\delta_{\ell,2}+
    \frac{1+(-1)^\ell(1 -4\nu^2)}{8}
    \, \frac{\Gamma(\ell)}{(4n)^\ell}\,\left(
        1-\nu \,\frac{\ell}{2n}
\right) + {\cal O}(n^{-(\ell+2)}).\nonumber\\{}
\end{eqnarray}
For $\ell$ kept finite, the next-order terms in $1/n$ can be derived with increasing difficulty.

\noindent\newline\newline
{\it (iii)} The $\{c_\ell\}$--sequence: For instance, one may show that
the term of the order ${\mathcal O}(n^{-(\ell+2)})$ in Eq.~(\ref{g-exp-1n}) equals
\begin{eqnarray}
             (-1)^\ell \Gamma(\ell) \,\frac{c_\ell(\nu)}{(4\nu)^{\ell+2}},
\end{eqnarray}
where
\begin{eqnarray}\label{c-addition}
    c_\ell(\nu) &=& \frac{\ell}{96} \Big[ \Big(
        12\nu^2 (2\ell+3) +3 (3\ell^2-4)\Big)\nonumber\\ &-& (-1)^\ell (1-4\nu^2) \Big( 4\nu^2 (\ell+1)(\ell-7) - 3
(3\ell^2-4)\Big) \Big].
\end{eqnarray}
Let us reiterate that the above $1/n$ expansion of $\kappa_\ell(G)$ refers to finite $\ell$.

Equations (\ref{g-exp-1n}) -- (\ref{c-addition}) extend earlier asymptotic results by Brouwer and Beenakker (1996), and Savin, Sommers and Wieczorek (2008) to the case of cumulants of arbitrary finite order. For an extension of the above analysis to other Dyson's symmetry classes ($\beta=1$ and $4$), the reader is referred to the recent study by Mezzadri and Simm (2013).

\section{Integrable theory of noise fluctuations in chaotic cavities with ideal leads}\label{itnf}
\subsection{Charge fluctuations and the noise power} \label{charge-f}
The charge transfer through a phase-coherent
cavity exhibiting chaotic classical dynamics is a random process
influenced by discreteness of the electron charge $e$ and the
quantum nature of electrons (Blanter and B\"uttiker 2000, Imry 2002). Fluctuations
of charge transmitted during a fixed time interval or, equivalently,
fluctuations $\delta I(t)$ of current around its mean are quantified
by the noise power (Lesovik 1989, B\"uttiker 1990)
\begin{eqnarray}
    {\mathcal P} = 2\int_{-\infty}^{+\infty} dt\, \left< \delta I(t+t_0) \delta I (t_0)
    \right>_{t_0},
\end{eqnarray}
where the brackets $\left<\cdots \right>_{t_0}$ indicate averaging over the reference time $t_0$.

At temperatures $\theta = k_B T$ which are much larger than a bias
voltage ${\upsilon}=eV$ applied to the cavity ($\theta \gg
\upsilon$), the current fluctuations are dominated by the
equilibrium {\it thermal noise}, also known as Johnson-Nyquist
noise. Caused by fluctuating occupation numbers in a flow of
carriers injected into cavity from electronic reservoirs, thermal
noise extends over all frequencies up to the quantum limit
$\theta/h$. In the absence of electron-electron interactions, its
power at zero bias voltage ($\upsilon =0$) is related to the
scattering matrix ${\bm {\mathcal S}}$ of the system composed of the
cavity and the leads (Khlus 1987, Lesovik 1989, B\"uttiker 1990,
Martin and Landauer 1992, B\"uttiker 1992):
\begin{eqnarray}
\label{pth}
    {\mathcal P}_{\rm th}(\theta) = 4\theta\,G_0\,
    {\rm tr} ({\mathcal C}_1 {\bm {\mathcal S}} {\mathcal C}_2 {\bm {\mathcal S}}^\dagger).
\end{eqnarray}
Here, $G_0= e^2/h$ is the conductance quantum. The projection
matrices ${\mathcal C}_{1,2}$ encoding the information about
particular cavity-lead geometry are specified in Eq.~(\ref{c1-c2}).

In the opposite limit of low temperatures ($\theta \ll \upsilon$),
the current fluctuations are still significant even though the flow
of incident electrons is essentially noiseless. In this temperature
regime, nonequilibrium current fluctuations (known as a {\it shot
noise}) exist because of (i) the granularity of the electron charge
$e$ and (ii) the stochastic nature of electron scattering inside the
cavity which splits the electron wave into two or more partial waves
leaving the cavity through different exits. It is this ``uncertainty
of not knowing where the electron came from and where it will go
to'' (Oberholzer {\it et al} 2002) that makes the transmitted charge
to fluctuate. At zero temperature, the scattering matrix approach
brings the shot noise power in the form
\begin{equation}
\label{pshot}
    {\mathcal P}_{\rm shot}(\upsilon) = 2 \upsilon\,G_0\left[
    {\rm tr} ({\mathcal C}_1 {\bm {\mathcal S}} {\mathcal C}_2 {\bm {\mathcal S}}^\dagger)
    - {\rm tr} ({\mathcal C}_1 {\bm {\mathcal S}} {\mathcal C}_2 {\bm {\mathcal S}}^\dagger)^2
    \right].
\end{equation}

At finite temperatures, both sources of noise are operative, the
total noise ${\mathcal P}(\theta,\upsilon)$ being a complicated
function of temperature and bias voltage
\footnote{Equation~(\ref{ptotal}) disregards the low-frequency $1/f$
noise that can efficiently be filtered out in experiments.}:
\begin{eqnarray}\fl
\label{ptotal}
    {\mathcal P}(\theta,\upsilon) = 4 \theta \, G_0\,
    \Big(
    {\rm tr} ({\mathcal C}_1 {\bm {\mathcal S}}
  {\mathcal C}_2 {\bm {\mathcal S}}^\dagger)^2
    + \frac{\upsilon}{2\theta} \, {\rm coth}
    \left(
        \frac{\upsilon}{2\theta}
    \right)
    \left[
    {\rm tr} ({\mathcal C}_1 {\bm {\mathcal S}} {\mathcal C}_2 {\bm {\mathcal S}}^\dagger)
    - {\rm tr} ({\mathcal C}_1 {\bm {\mathcal S}} {\mathcal C}_2 {\bm {\mathcal S}}^\dagger)^2
    \right] \Big).
\end{eqnarray}
Equation (\ref{ptotal}) suggests that the crossover from thermal
noise ${\mathcal P}_{\rm th}(\theta)= {\mathcal P}(\theta,0)$ to
shot noise ${\mathcal P}_{\rm shot}(\upsilon)= {\mathcal
P}(0,\upsilon)$ depends in a sensitive way on scattering properties
of the cavity and the leads incorporated in the scattering matrix
${\bm {\mathcal S}}$. Since chaotic scattering of electrons inside the
cavity induces fluctuations of ${\bm {\mathcal S}}$-matrix,
the noise power ${\mathcal P}(\theta,\upsilon)$ fluctuates, too.

So far, the thermal to shot noise crossover has only been studied at
the level of {\it average} noise power. For the two-terminal
scattering geometry comprised of the cavity attached to outside
reservoirs (kept at temperature $\theta$) via two leads supporting
$N_{\rm L}$ and $N_{\rm R}$ propagating modes, respectively, the
average noise power equals (Blanter and Sukhorukov 2000, Oberholzer
{\it et al} 2001, Savin and Sommers 2006)
\begin{equation}
\label{paverage}
 \left<{\mathcal P}(\theta,\upsilon)\right>_{{\bm {\mathcal S}}} =
 \left<{\mathcal P}_{\rm th}\right>_{{\bm {\mathcal S}}}
 \left[
        1 + \frac{N_{\rm L}N_{\rm R}}{(N_{\rm L}+N_{\rm R})^2-1}\, f_\eta
 \right],
\end{equation}
where
\begin{equation}
\label{peq}
 \left<{\mathcal P}_{\rm th}\right>_{{\mathcal S}} =
 4\theta\, G_0 \frac{N_{\rm L} N_{\rm R}}{N_{\rm L}+N_{\rm R}}
\end{equation}
is the average equilibrium thermal noise power, and the
thermodynamic function
\begin{eqnarray}
\label{fbeta}
    f_\eta = \eta \coth \,\eta -1
\end{eqnarray}
depends on the ratio
\begin{eqnarray} \label{beta-def}
\eta=\frac{\upsilon}{2\theta}
\end{eqnarray}
between the bias
voltage $\upsilon$ and the temperature $\theta$. Equations
(\ref{paverage}) and (\ref{peq}) hold for cavities with broken time
reversal symmetry; the two can readily be extended to other
symmetry classes (Savin and Sommers 2006). Derived for the
universal transport regime (Beenakker 1997, Richter and Sieber 2002,
M\"uller {\it et al} 2007) emerging in the limit $\tau_{\rm D} \gg
\tau_{\rm E}$ (Agam {\it et al} 2000), where $\tau_{\rm D}$ is the
average electron dwell time and $\tau_{\rm E}$ is the Ehrenfest time
(the time scale where quantum effects set in), the above prediction
has been confirmed in a remarkable series of experiments (Oberholzer
{\it et al} 2001, Cron{\it et al} 2001, Oberholzer {\it et al}
2002).

Below, we examine {\it statistics} of the thermal to shot
noise crossover. The latter, contained in the distribution function
of the noise power ${\mathcal P}(\theta,\upsilon)$ or, equivalently,
in its {\it cumulants}, can effectively be described within the framework
of the formalism exposed in Section 2.

\subsection{Statistics of shot noise fluctuations and Painlev\'e V (easy way)}\noindent
{\it Symmetric leads.}---To start with, we shall attempt extending an `easy way' approach outlined in Section 2 to nonperturbatively describe statistics of noise fluctuations at $T=0$ when the thermal noise contribution vanishes. According to Eq. (\ref{pshot}), taken in conjunction with reflection-transmission-decomposed scattering matrix [Eq. (\ref{s-block})], the MGF
\begin{eqnarray}
\tilde{{\mathcal F}}_n^{(\nu)}(z) = \left<
    \exp \left(z\, P_{\rm shot}\right)
    \right>_{\bm{\mathcal{S}} \in {\rm CUE}(N_{\rm L}+N_{\rm R})}
\end{eqnarray}
for dimensionless shot-noise power
\begin{eqnarray}
    P_{\rm shot}({\bm T}) = (2\upsilon G_0)^{-1} {\mathcal P}_{\rm shot}(\upsilon) = \sum_{j=1}^n T_j (1-T_j)
\end{eqnarray}
takes the form
\begin{eqnarray}\label{sn-mgf}
    \tilde{{\mathcal F}}_n^{(\nu)}(z) =     c_{n,\nu}^{-1} \int_{(0,1)^n} \prod_{j=1}^n dT_j \, T_j^\nu \exp \left(
    z\, T_j(1-T_j)
    \right) \cdot \Delta_n^2({\bm T}).
\end{eqnarray}
Similarly to the MGF for Landauer conductance, we make use of the Andr\'eief--de Bruijn integration formula to reduce $\tilde{{\mathcal F}}_n^{(\nu)}(z)$ to the Hankel determinant
\begin{eqnarray}\hspace{-1cm}
    \tilde{{\mathcal F}}_n^{(\nu)}(z) = \frac{n!}{c_{n,\nu}} \,{\rm det}
    \left[
    \int_0^1 dT \, T^{\nu+j+k} \exp \left(
    z\, T(1-T)
    \right)\right]_{(j,k)\in (0,n-1)}
\end{eqnarray}
or, equivalently,
\begin{eqnarray}
\label{hd-2}\hspace{-1.5cm}
    \tilde{{\mathcal F}}_n^{(\nu)}(z) = \frac{n!}{c_{n,\nu}} \,e^{nz/4} {\rm det}
    \left[
    \int_{-1/2}^{+1/2} dx \, \left(x+\frac{1}{2}\right)^{\nu} \, x^{j+k} e^{-z x^2}
    \right]_{(j,k)\in (0,n-1)}.
\end{eqnarray}
Unfortunately, for a generic $\nu$, there is no obvious way to represent the $(j,k)$ entry in Eq.~(\ref{hd-2}) in an operator form similar to the one appearing in Eq.~(\ref{hd}). However, for the case of {\it symmetric} leads ($\nu=0$), one can readily spot that $\tilde{{\mathcal F}}_n^{(0)}(z)$ reduces to the product of two Hankel determinants, each of them possessing the desired operator structure:
\begin{eqnarray}
\label{hd-f}
    \tilde{{\mathcal F}}_n^{(0)}(z) =
        e^{n z/4}    {\tilde {\mathcal F}}_{\lfloor n/2 \rfloor}^{\,+} (z)
                {\tilde {\mathcal F}}_{\lceil n/2 \rceil}^{\,-} (z),
\end{eqnarray}
where
\begin{eqnarray}\hspace{-1cm}
        {\tilde {\mathcal F}}_\ell^{\,\pm} (z) =
        c_{\ell,\pm 1/2}^{-1} \, \ell!\, {\rm det} \left[ \int_0^1
                d\lambda \, \lambda^{\pm 1/2} e^{-(z/4)\lambda} \lambda^{j+k}
                \right]_{(j,k)\in (0,\ell-1)}.
\end{eqnarray}
The above factorisation is a consequence of the checkerboard structure of the moment matrix in Eq.~(\ref{hd-2}) at $\nu=0$,
\begin{eqnarray}
\int_{-1/2}^{+1/2} dx  \, x^{j+k} e^{-z x^2} \propto \delta_{j+k,{\rm even}}.
\end{eqnarray}
In a backward move, we have
\begin{eqnarray}
        {\tilde {\mathcal F}}_\ell^{\,\pm} (z) =
        c_{\ell,\pm 1/2}^{-1} \int_{(0,1)^\ell} \prod_{j=1}^\ell
                d\lambda_j \, \lambda_j^{\pm 1/2} e^{-(z/4)\lambda_j} \cdot \Delta_\ell^2({\bm \lambda}).
\end{eqnarray}
The latter, being essentially a gap formation probability in the LUE ensemble, can conveniently be expressed in terms of the Landauer conductance MGF [see Eq.~(\ref{cond-eig-nod})] taken at effective asymmetry parameter $\nu = \pm 1/2$,
\begin{eqnarray}
        {\tilde {\mathcal F}}_\ell^{\,\pm} (z) = {\mathcal F}_\ell^{(\pm 1/2)} \left(\frac{z}{4}\right).
\end{eqnarray}
As the result, we observe the relation
\begin{eqnarray}
\label{hd-ff}
    \tilde{{\mathcal F}}_n^{(0)}(z) =\,
        e^{n z/4}   {\mathcal F}_{\lfloor n/2 \rfloor}^{(+1/2)} \left(\frac{z}{4}\right)
                {\mathcal F}_{\lceil n/2 \rceil}^{(-1/2)} \left(\frac{z}{4}\right)
\end{eqnarray}
which, if taken together with the Painlev\'e V representation of ${\mathcal F}_n^{(\nu)}(z)$ [Eqs.~(\ref{FnP-22}), (\ref{pv-22}) and (\ref{pv-bc})], yields the shot noise MGF in the form
\begin{eqnarray}
\label{FnP-33}\hspace{-1.5cm}
    \tilde{{\mathcal F}}_n^{(0)}(z) = \, e^{n z/4} \exp\left(
        \int_0^{z/4} dt \frac{\sigma_{\lfloor n/2 \rfloor,+1/2}(t) + \sigma_{\lceil n/2 \rceil,-1/2}(t) - n^2/2}{t}
    \right).
\end{eqnarray}
We remind that the above result holds for the symmetric chaotic cavity probed via symmetric leads, $N_{\rm L} = N_{\rm R}=n$. Owing to the cumulant expansion
\begin{eqnarray}
   \log \tilde{{\mathcal F}}_n^{(0)}(z) = \sum_{\ell=1}^\infty \frac{z^\ell}{\ell!}\, \kappa_\ell^{(n,0)}(P_{\rm shot}),
\end{eqnarray}
we are now able to express the shot noise cumulants $\kappa_\ell^{(n,0)}(P_{\rm shot})$ through ``effective cumulants'' $\kappa_\ell ^{(n,\nu)}(G)$ of the Landauer conductance (with $\nu = \pm 1/2$)
defined by the Taylor expansion Eq.~(\ref{main-cond-cum}). Straightforward calculation brings
\begin{eqnarray}\fl
    \kappa_\ell^{(n,0)}(P_{\rm shot}) = \frac{n}{4}\delta_{\ell,1}+\frac{(-1)^\ell}{4^\ell} \left(
    \kappa_\ell^{(\lceil n/2 \rceil,-1/2)}(G) + \kappa_\ell^{(\lfloor n/2 \rfloor,+1/2)}(G)
    \right),
\end{eqnarray}
where $\kappa_\ell^{(m,\pm 1/2)}(G)$  are the solutions to the recurrence equation Eq.~(\ref{cumeq}) taken at $n=m$ and $\nu = \pm 1/2$, and supplemented by appropriate initial conditions [Eq.~(\ref{ic-cond})]. In particular, one has:
\begin{eqnarray} \label{psh-1}
    \kappa_1^{(n,0)}(P_{\rm shot}) &=& \frac{n^3}{2(4n^2-1)}, \\
    \label{psh-2}
    \kappa_2^{(n,0)}(P_{\rm shot}) &=& \frac{n^2 (4n^4 -9n^2+3)}{8(4 n^2-1)^2 (4n^2-9)},\\
    \label{psh-3}
    \kappa_3^{(n,0)}(P_{\rm shot}) &=& \frac{n^2 (16 n^6 -24 n^4+9 n^2 +1)}{128(4 n^2-1)^4}.
\end{eqnarray}
Whenever an overlap exists, Eqs.~(\ref{psh-1}) -- (\ref{psh-3}) coincide with the earlier results by Savin and Sommers (2006) and Savin, Sommers and Wieczorek (2008). Nonperturbative formulae for higher order cumulants of the shot noise can readily be generated as well, even though the corresponding expressions become increasingly cumbersome.
\noindent\newline\newline
{\it Asymmetric leads.}---Unfortunately, a transparent calculational framework outlined above fails to accommodate statistics of shot noise fluctuations in the cavities probed via {\it asymmetric} leads when the factorisation Eq. (\ref{hd-f}) is no longer available. Also, an attempt to formulate a $\tau$ function theory [see Section 2.3] of the eigenvalue integral Eq. (\ref{sn-mgf}) considered at $\nu \neq 0$ brings no fruit: projecting the first KP equation for the associate $\tau$ function onto the hyperplane ${\bm t}={\bm 0}$ with the help of Virasoro constraints turns out to be a nontrivial task since an {\it infinitely many} Virasoro constraints appear to be relevant. In the forthcoming Section, we show how this technical obstacle can be overcome by considering a joint distribution of the Landauer conductance and the noise power.

\subsection{Joint moment generating function of Landauer conductance and the noise power}
A way out of this difficulty was reported by Osipov and Kanzieper (2009) where it was demonstrated that the shot noise moment generating function can be restored from a two-dimensional lattice equation for the {\it joint} moment generating function (JMGF) of the Landauer conductance and the shot noise power. To make a contact with experiments, below we consider a JMGF of the Landauer conductance and the noise power ${\mathcal P}(\theta,\upsilon)$ consisting of two competing contributions -- the shot and the thermal noise (see Section \ref{charge-f}).

The starting point of our analysis is the JMGF
\begin{eqnarray}
\label{cgf-def}
    {\mathcal F}_{n}^{(\nu)}(z,w) = \left<
        \exp(-z \,G)\exp(- w\, P)
    \right>_{{\bm {\mathcal S}}\in {\rm CUE}(N_{\rm L}+N_{\rm R})}
\end{eqnarray}
of the dimensionless Landauer conductance $G = {\rm tr} ({\mathcal C}_1
{\bm {\mathcal S}} {\mathcal C}_2 {\bm {\mathcal S}}^\dagger)$ and the dimensionless noise
power
\begin{eqnarray}
    P =
    {\rm tr} ({\mathcal C}_1 {\bm {\mathcal S}}
  {\mathcal C}_2 {\bm {\mathcal S}}^\dagger)^2
    + \eta \, {\rm coth}\, \eta\,
    \left[
    {\rm tr} ({\mathcal C}_1 {\bm {\mathcal S}} {\mathcal C}_2 {\bm {\mathcal S}}^\dagger)
    - {\rm tr} ({\mathcal C}_1 {\bm {\mathcal S}} {\mathcal C}_2 {\bm {\mathcal S}}^\dagger)^2
    \right]
\end{eqnarray}
that corresponds to the noise power ${\mathcal P}(\theta,\upsilon)$ measured in the units $4\theta G_0$.
The notation
${\bm {\mathcal S}} \in {\rm CUE}(N_{\rm L}+N_{\rm R})$ indicates that averaging runs over
scattering matrices ${\bm {\mathcal S}}$ drawn from the Dyson circular
unitary ensemble as discussed in Section \ref{intro}. The
joint dimensionless cumulants
\begin{eqnarray}\label{gp-cum}
\kappa_{\ell,m}^{(n,\nu)}(G,P) =
    \langle\!\langle G^\ell P^m
    \rangle\!\rangle
\end{eqnarray}
can be extracted from the expansion
\begin{equation}
\label{c-def}
    \log {\mathcal F}_n^{(\nu)}(z,w) = \sum_{\ell,m=0}^\infty (-1)^{\ell+m} \frac{z^\ell w^m}{\ell!\,m!}\,
    \kappa_{\ell,m}^{(n,\nu)}(G,P),
\end{equation}
where $\kappa_{0,0}\equiv 0$ and the subscript $n$ stands for $n=\min(N_{\rm
L},N_{\rm R})$. We notice in passing that a somewhat simpler joint cumulant of Landauer conductance
and the {\it shot} noise power
\begin{eqnarray}
\tilde{\kappa}_{\ell,m}^{(n,\nu)}(G,P_{\rm shot}) =
    \langle\!\langle G^\ell P_{\rm shot}^m
    \rangle\!\rangle
\end{eqnarray}
frequently considered in the literature can be obtained from Eq.~(\ref{gp-cum}) via the limiting procedure
\begin{eqnarray}\label{lim-rel}
    \tilde{\kappa}_{\ell,m}^{(n,\nu)}(G,P_{\rm shot}) = \lim_{\eta\rightarrow \infty} \frac{1}{\eta^m} \, \kappa_{\ell,m}^{(n,\nu)}(G,P).
\end{eqnarray}
To perform the averaging in Eq.~(\ref{cgf-def}) in a most economic
way, we employ a decomposition Eq.~(\ref{s-block}) to bring into play a
set of $n$ transmission eigenvalues ${\bm T} = (T_1,\cdots,T_n) \in
(0,1)^n$ distributed in accordance with the joint probability
density function given by Eq. (\ref{PnT}) and consequently derive the JMGF in the form \footnote[5]{The same eigenvalue integral appears in replica calculations of {\it dynamic} correlation functions in the Calogero-Sutherland model with the interaction parameter $\lambda=1$, see: Gangardt and Kamenev (2001).}:
\begin{equation}
\label{Fnzw}
    {\mathcal F}_n^{(\nu)}(z,w) = c_{n,\nu}^{-1}\int_{(0,1)^n} \prod_{j=1}^n dT_j\,T_j^\nu \,\Gamma_{z,w}(T_j)
    \,\Delta_n^2({\bm T}),
\end{equation}
where
\begin{eqnarray}
\label{Gamma}
    \Gamma_{z,w}(T) =
    \exp\left[
        -(z+w)\,T - w\,f_\eta\, T(1-T)
    \right].
\end{eqnarray}
Here, the exponent contains weighted contributions from the conductance
$G({\bm T}) = \sum_{j=1}^n T_j$ and the noise power
\begin{eqnarray}
\label{ptotal-T}
    P({\bm T})
    =
    \sum_{j=1}^n T_j + f_\eta \sum_{j=1}^n T_j(1-T_j).
\end{eqnarray}
The thermodynamic function $f_\eta$ is defined by Eq.~(\ref{fbeta}). We also recall that parameter $\nu$ in Eq.~(\ref{Fnzw}) is a measure of asymmetry
between the leads [Eq.~(\ref{nu-index})], the
notation $\Delta_n({\bm T})$ stands for the Vandermonde determinant
$\Delta_n ({\bm T})  =\prod_{j<k} (T_k-T_j)$, whilst $c_{n,\nu}$ is a
normalisation constant specified by Eq. (\ref{nc}).

Although the above matrix integral representation of the JMGF
${\mathcal F}_n^{(\nu)}(z,w)$ is by far more complicated than the one
appearing in the integrable theory of conductance fluctuations
(see Section 2),
\begin{eqnarray} \fl \label{ccgf}
    {\mathcal F}_n^{(\nu)}(z,0) =
    \left<
        \exp(-z \,G)\right>_{{\bm {\mathcal S}}\in {\rm CUE}(N_{\rm L}+N_{\rm R})}
        =
        c_{n,\nu}^{-1}\int_{(0,1)^n} \prod_{j=1}^n dT_j\,T_j^\nu \,e^{-z T_j}
    \,\Delta_n^2({\bm T}), \nonumber\\
    &&
\end{eqnarray}
it can still be treated nonperturbatively.

\subsection{The $\tau$ function theory of the joint moment generating function}

\subsubsection{The KP equation and Virasoro constraints}\label{s-223a}\noindent\newline\newline
To determine ${\mathcal F}_n^{(\nu)}(z,w)$ nonperturbatively, we define the $\tau$ function
\begin{equation}
    \label{tau-def}
    \tau_n^{(\nu)}({\bm t};z,w) = \frac{1}{n!}\int_{(0,1)^n} \prod_{j=1}^n dT_j\,T_j^\nu \,\Gamma_{z,w}(T_j)
    \,e^{V({\bm t};T_j)} \Delta_n^2({\bm T})
\end{equation}
whose evolution is governed by the bilinear identity Eq.~(\ref{bi-id}). The latter generates a wealth of nonlinear
differential hierarchical relations between various $\tau$ functions in the $(n, {\bm t},z,w)$ space. Of those, we are interested in the first KP equation
Eq.~(\ref{fkp}) because its projection onto the hyperplane ${\bm t}={\bm 0}$,
\begin{eqnarray}
\label{proj}
    {\mathcal F}^{(\nu)}_n(z,w) = \frac{n!}{c_{n,\nu}}\,
    \tau_n^{(\nu)}({\bm t};z,w)\Big|_{{\bm t}={\bm 0}},
\end{eqnarray}
will generate a closed nonlinear differential
equation for the JMGF ${\mathcal F}_n^{(\nu)}(z,w)$. It is this equation that will further be utilized to determine the noise power cumulants we are aimed at.

Since we are interested in deriving a differential equation for
${\mathcal F}_n^{(\nu)}(z,w)$ in terms of the derivatives with respect to variables $z$
and $w$, we have to use the {\it Virasoro constraints}, much in line with our treatment of conductance cumulants in Section 2. Demanding the invariance of $\tau_n^{(\nu)}({\bm t};z,w)$ under
the same set of transformations Eq.~(\ref{v-tr}), one readily derives the following set of Virasoro
constraints:
\begin{equation}
\label{vc-1}
    \big[ \hat{L}_{q+1}({\bm t}) - \hat{L}_{q}({\bm t})\big] \tau_n^{(\nu)}({\bm t};z,w)
     = 0,\;\;\;  q\ge 0,
\end{equation}
where
\begin{eqnarray}\fl
\label{vc-2}
    \hat{L}_q({\bm t})
     =
     \hat{\mathcal L}_{q}({\bm t}) + 2f_\eta\, w
        \frac{\partial}{\partial t_{q+2}} - [z+(1+f_\eta)\,w] \frac{\partial}{\partial t_{q+1}}
        + \nu \frac{\partial}{\partial t_q}
\end{eqnarray}
involves the Virasoro operators Eq.~(\ref{vc-3}) satisfying the Virasoro algebra Eq.~(\ref{va}). The usual convention
$\partial/\partial t_0 \equiv n$ is everywhere assumed.

\subsubsection{Joint MGF and the Toda Lattice equation}\label{TL-223}\noindent\newline\newline
Similarly to the MGF for Landauer conductance, the joint MGF ${\mathcal F}_n^{(\nu)}(z,w)$ can be shown to satisfy an one-dimensional
Toda Lattice equation. For one, it can be derived from the first equation of the TL hierarchy [Eq.~(\ref{ftl})], the identity Eq.~(\ref{id-11}) below, and
the projection relation Eq.~(\ref{proj}),
\begin{equation}
\label{TL-GP}\fl
    {\mathcal F}_n^{(\nu)}(z,w)\,\frac{\partial^2 {\mathcal F}_n^{(\nu)}(z,w)}{\partial z^2} - \left(
    \frac{\partial {\mathcal F}_n^{(\nu)}(z,w)}{\partial z}\right)^2 = {\rm var}_{n,\nu}(G)\,
    {\mathcal F}_{n-1}^{(\nu)}(z,w)\, {\mathcal F}_{n+1}^{(\nu)}(z,w),
\end{equation}
where ${\rm var}_{n,\nu}(G)$ is the conductance variance. Initial conditions read:
\begin{eqnarray}
    {\mathcal F}_0^{(\nu)}(z,w) = 1
\end{eqnarray}
and
\begin{eqnarray}
    {\mathcal F}_1^{(\nu)}(z,w) = (\nu+1) \int_0^1 dT\, T^\nu e^{-(z+w)T - w f_\eta T(1-T)}.
\end{eqnarray}
Iterating the TL equation $n$ times and calculating a double inverse Laplace transform, one may in principle restore a joint distribution function of the conductance and the the noise power. However, such a procedure is by far more complicated technically as compared to the conductance distribution (see Section \ref{TL-Hankel}).

\subsubsection{Nonlinear differential equation for the joint MGF}\label{s-223}\noindent\newline\newline
Aiming at deriving a differential equation for the JMGF ${\mathcal
F}_n^{(\nu)}(z,w)$, we project the first KP equation Eq.~(\ref{fkp})
onto the hyperplane ${\bm t}={\bm 0}$. Spotting the identities
($f_\eta>0$)
\begin{eqnarray}\label{id-11}
\frac{\partial}{\partial t_1} \tau_n^{(\nu)}({\bm t};z,w)&=& - \frac{\partial}{\partial z}  \tau_n^{(\nu)}({\bm t};z,w),\\
f_\eta \frac{\partial}{\partial t_2}  \tau_n^{(\nu)}({\bm t};z,w) &=& \frac{\partial}{\partial w}  \tau_n^{(\nu)}({\bm t};z,w)
- (1+f_\eta)\frac{\partial}
{\partial z}  \tau_n^{(\nu)}({\bm t};z,w),
\end{eqnarray}
we combine Eqs.~(\ref{fkp}) with the Virasoro constraint
Eq.~(\ref{vc-1}) taken at $q=0$,
\begin{eqnarray} \fl
    \Bigg[ \sum_{j=1}^\infty jt_j \left(
    \frac{\partial}{\partial t_{j+1}}
    -
    \frac{\partial}{\partial t_{j}}
    \right)
    +  2f_\eta\, w \left(
        \frac{\partial}{\partial t_{3}} -
        \frac{\partial}{\partial t_{2}}
    \right) \nonumber \\ \fl
         \quad - \left[z+(1+f_\eta)\,w\right] \left(
     \frac{\partial}{\partial t_{2}}
     -
     \frac{\partial}{\partial t_{1}}
     \right)
     + (2n+\nu) \frac{\partial}{\partial t_1}
     - n(n+\nu) \Bigg]\,  \tau_n^{(\nu)}({\bm t};z,w)=0,
\end{eqnarray}
to derive:
\begin{eqnarray}\fl
\label{ndeq}
    \Bigg[ w f_\eta^2 \frac{\partial^4}{\partial z^4} + \left[2(2n+\nu)f_\eta -  2 z + w\,(1-f_\eta^2)\right]
    \frac{\partial^2}{\partial z^2} + 2(z-2w)\frac{\partial^2}{\partial z \partial w}
    \nonumber \\  \fl
     \quad + 3 w \frac{\partial^2}{\partial w^2} + 2 \left(\frac{\partial}{\partial w} -  \frac{\partial}{\partial z}\right) \Bigg]
    \,\log {\mathcal F}_n^{(\nu)}(z,w)+ 6 w\,f_\eta^2  \left(
        \frac{\partial^2}{\partial z^2} \, \log {\mathcal F}_n^{(\nu)}(z,w)
    \right)^2
      = 0. \nonumber\\
      {}
\end{eqnarray}
Owing to Eq.~(\ref{c-def}), the nonlinear equation Eq.~(\ref{ndeq}) supplemented by the initial condition
\begin{eqnarray}
    {\mathcal F}_n^{(\nu)}(z,w=0) = \exp\left(
        \int_0^z dt \, \frac{\sigma_{n,\nu}(t) -n(n+\nu)}{t}
    \right),
\end{eqnarray}
see Eqs.~(\ref{sF-cc}) and (\ref{OK-2008}) [or (\ref{pv-22})], contains all
the information about joint cumulants of the Landauer conductance
and the noise power. Although finding explicit formulae expressing ${\mathcal F}_n^{(\nu)}(z,w)$ in terms of the fifth Painlev\'e transcendent $\sigma_{n,\nu}$ is an unviable task, the joint cumulants $\kappa_{\ell,m}^{(n,\nu)}(G,P)$ can nonetheless be related to the cumulants of Landauer conductance $\kappa_\ell^{(n,\nu)}(G)$. This will be done in the next Section.

\subsection{Joint cumulants of Landauer conductance and the noise power}
\subsubsection{Exact recurrence solution for joint cumulants}\noindent\newline\newline
Indeed, combining the cumulant expansion Eq.~(\ref{c-def}) with the
differential equation Eq.~(\ref{ndeq}), we deduce after some algebra a nonlinear
recurrence for the joint dimensionless cumulants of conductance and noise power ($\ell,m \ge 0$):
\begin{eqnarray}
\label{2drec} \fl
    m\,\Big[f_\eta^2\, \kappa_{\ell+4,m-1}^{(n,\nu)}(G,P) + (1-f_\eta^2) \,\kappa_{\ell+2,m-1}^{(n,\nu)}(G,P)\Big] - 2 \,
    (2n+\nu)\,  f_\eta \,\kappa_{\ell+2,m}^{(n,\nu)}(G,P) \nonumber \\
    \fl \qquad\quad
    - 2\left(\ell+ 2 m + 1 \right) \, \kappa_{\ell+1,m}^{(n,\nu)}(G,P)    +  (2\ell+3m+2) \, \kappa_{\ell,m+1}^{(n,\nu)}(G,P)
    \nonumber\\
    \fl \qquad\quad
     + \, 6m \,f_\eta^2
    \sum_{i=0}^{m-1}\left( {m-1}\atop{i} \right) \sum_{j=0}^\ell   \left( {\ell}\atop{j} \right)
    \kappa_{j+2,i}^{(n,\nu)}(G,P)\, \kappa_{\ell-j+2,m-i-1}^{(n,\nu)}(G,P)=0. \nonumber\\
    {}
\end{eqnarray}
By virtue of the limiting relation Eq.~(\ref{lim-rel}), one readily generates a nonlinear recurrence for zero-temperature
joint cumulants of Landauer conductance and the shot noise power:
\begin{eqnarray}
\label{2drec-shot} \fl
    m\,\Big[\tilde{\kappa}_{\ell+4,m-1}^{(n,\nu)}(G,P_{\rm shot}) - \tilde{\kappa}_{\ell+2,m-1}^{(n,\nu)}(G,P_{\rm shot})\Big] - 2 \,
    (2n+\nu)\, \tilde{\kappa}_{\ell+2,m}^{(n,\nu)}(G,P_{\rm shot}) \nonumber \\
    \fl \qquad\quad
    +  (2\ell+3m+2) \, \tilde{\kappa}_{\ell,m+1}^{(n,\nu)}(G,P_{\rm shot})
    \nonumber\\
    \fl \qquad\quad
     + \, 6m \,
    \sum_{i=0}^{m-1}\left( {m-1}\atop{i} \right) \sum_{j=0}^\ell   \left( {\ell}\atop{j} \right)
    \tilde{\kappa}_{j+2,i}^{(n,\nu)}(G,P_{\rm shot})\, \tilde{\kappa}_{\ell-j+2,m-i-1}^{(n,\nu)}(G,P_{\rm shot})=0. \nonumber\\
    {}
\end{eqnarray}
To resolve these two-dimensional recursion equations, one must know the cumulants $\kappa_{\ell,0}^{(n,\nu)} = \tilde{\kappa}_{\ell,0}^{(n,\nu)}= \kappa_\ell^{(n,\nu)}(G)$ of dimensionless
conductance which play a r\^ole of boundary conditions. Since these are
known [see Eqs.~(\ref{cumeq})], Eqs. (\ref{2drec}) and (\ref{2drec-shot}) provide a nonperturbative solution to the problem
of noise power fluctuations by relating the latter to the cumulants
of Landauer conductance. [For Dyson's symmetry indices $\beta=1$ and $4$, an analogue of Eq.~(\ref{2drec-shot}) was recently derived by Mezzadri and Simm (2013).]

Undoubtedly, the very existence of the above nontrivial relation
(which emphasises a fundamental r\^{o}le played by Landauer
conductance in transport problems) must be well rooted in the
mathematical formalism and also have a good physics reason. As far
as the former point is concerned, we wish to stress that a na\"{i}ve
attempt to build a theory for the generating function ${\mathcal
F}_n^{(\nu)}(0,w)$ of solely noise power cumulants faces an unsurmountable
obstacle: the KP equation [Eq.~(\ref{fkp})] and appropriate Virasoro
constraints [Eq.~(\ref{vc-1}) at $z=0$] cannot be resolved jointly
in the hyperplane ${\bm t}={\bm 0}$. This justifies the starting
point [Eq.~(\ref{cgf-def})] of our analysis. The physics arguments
behind the peculiar structure of our solution are yet to be found.

\subsubsection{Noise power cumulants}\noindent\newline\newline
Some computational effort is required to read off explicit formulae
for the noise power cumulants from Eqs.~(\ref{2drec}) and
(\ref{cumeq}). Below we provide expressions for two families of
joint cumulants expressed in terms of dimensionless cumulants
$\kappa_{\ell}^{(n,\nu)}(G) =\langle \! \langle G^\ell \rangle\!\rangle$
of the Landauer conductance.
\noindent\newline\newline
{\it (i-a)} Mean noise power is generated by the lowest order member $(\ell,m)=(0,0)$
        of the recurrence Eq.~(\ref{2drec}); it reads:
\begin{eqnarray} \label{mnp}
        \kappa_1^{(n,\nu)}(P) = (2n+\nu) f_\eta\,
        \kappa_{2}^{(n,\nu)}(G) + \kappa_{1}^{(n,\nu)}(G).
\end{eqnarray}
Being in concert
        with the known expression Eqs. (\ref{paverage}) and
        (\ref{peq}), this result is a particular case of a more
        general formula
        \begin{eqnarray}
        \label{k-l-1}
            \kappa_{\ell,1}^{(n,\nu)}(G,P) = \kappa_{\ell+1}^{(n,\nu)}(G) + (2n+\nu)\frac{f_\eta}{\ell+1}\, \kappa_{\ell+2}^{(n,\nu)}(G).
        \end{eqnarray}
\noindent\newline
{\it (i-b)} Mean shot noise power can be derived by virtue of Eq.~(\ref{lim-rel}); we deduce:
\begin{eqnarray} \label{mnp-shot}
        \kappa_1^{(n,\nu)}(P_{\rm shot}) = (2n+\nu)\, \kappa_{2}^{(n,\nu)}(G) = \frac{n^2 (n+\nu)^2}{(2n+\nu)
    [(2n+\nu)^2-1]}
\end{eqnarray}
and
        \begin{eqnarray}
        \label{k-l-1-shot}
            \kappa_{\ell,1}^{(n,\nu)}(G,P_{\rm shot}) = \frac{2n+\nu}{\ell+1}\, \kappa_{\ell+2}^{(n,\nu)}(G).
        \end{eqnarray}
Compare Eq.~(\ref{mnp-shot}) with Eq.~(\ref{psh-1}).
\noindent\newline\newline
{\it (ii-a)} Noise power variance is supplied by the $(\ell,m)=(0,1)$ member of the recurrence,
        \begin{eqnarray}
 \label{p-cum-2}\fl
 \kappa_2^{(n,\nu)}(P) =
    \left(
        \frac{2}{3}(2n+\nu)^2 -1
    \right) \frac{f_\eta^2}{5} \,\kappa_{4}^{(n,\nu)}(G) + (2n+\nu) f_\eta\, \kappa_{3}^{(n,\nu)}(G)
    \nonumber\\
    \qquad \qquad +\left(
        1+ \frac{f_\eta^2}{5}
    \right)\, \kappa_{2}^{(n,\nu)}(G) - \frac{6}{5} f_\eta^2 [\kappa_{2}^{(n,\nu)}(G)]^2.
\end{eqnarray}
Its generalisation reads:
\begin{eqnarray} \fl \label{k-l-2}
    \kappa_{\ell,2}^{(n,\nu)}(G,P) = \left(
        \frac{2(2n+\nu)^2}{\ell+3}-1
    \right) \frac{f_\eta^2}{2\ell+5}\kappa_{\ell+4}^{(n,\nu)}(G) + 2(2n+\nu) \frac{f_\eta}{\ell+2}\kappa_{\ell+3}^{(n,\nu)}(G)
    \nonumber\\
    +\left(
        1+ \frac{f_\eta^2}{2\ell+5}
    \right) \kappa_{\ell+2}^{(n,\nu)}(G)
    - 6 \frac{f_\eta^2}{2\ell+5} \sum_{j=0}^\ell \left({\ell}\atop{j}\right) \kappa_{j+2}^{(n,\nu)}(G)\,\kappa_{\ell+2-j}^{(n,\nu)}(G).
    \nonumber\\
    {}
\end{eqnarray}
\noindent\newline
{\it (ii-b)} Shot noise power variance, derived from Eqs.~(\ref{p-cum-2}) and (\ref{lim-rel}), equals
        \begin{eqnarray}
 \label{p-cum-2-shot}\fl
 \qquad
 \kappa_2^{(n,\nu)}(P_{\rm shot}) =
    \frac{1}{5}\Bigg[
    \left(
        \frac{2}{3}(2n+\nu)^2 -1
    \right)  \,\kappa_{4}^{(n,\nu)}(G)
    + \kappa_{2}^{(n,\nu)}(G) - 6 [\kappa_{2}^{(n,\nu)}(G)]^2\Bigg]. \nonumber\\
    {}
\end{eqnarray}
Similarly,
\begin{eqnarray} \fl \label{k-l-2-shot}
    \kappa_{\ell,2}^{(n,\nu)}(G,P_{\rm shot}) = \frac{1}{2\ell+5} \Bigg[\left(
        \frac{2(2n+\nu)^2}{\ell+3}-1
    \right) \,\kappa_{\ell+4}^{(n,\nu)}(G)  +
     \kappa_{\ell+2}^{(n,\nu)}(G) \nonumber\\
    \qquad \qquad \qquad \qquad - 6 \sum_{j=0}^\ell \left({\ell}\atop{j}\right) \kappa_{j+2}^{(n,\nu)}(G)\,\kappa_{\ell+2-j}^{(n,\nu)}(G)\Bigg].
\end{eqnarray}
Compare Eq.~(\ref{p-cum-2-shot}) with Eq.~(\ref{psh-2}).

The (joint) cumulants $\kappa_{\ell,m}^{(n,\nu)}(G,P)$ of higher order ($m \ge 3$)
can be calculated in the same manner albeit explicit expressions
become increasingly cumbersome.

\subsubsection{Asymptotic analysis of joint cumulants: Symmetric leads}\noindent\newline\newline
The nonperturbative solution Eq.~(\ref{2drec}) has
a drawback: it does not supply much desired {\it explicit}
formula of joint conductance and noise power cumulants
$\kappa_{\ell,m}^{(n,\nu)}(G,P)$. To probe the latter, we turn to the large-$n$
limit of the recurrence Eq.~(\ref{2drec}). In what follows, the
asymmetry parameter $\nu$ will be set to zero when the joint cumulants $\kappa_{\ell,m}^{(n,0)}(G,P)$ are solutions to the recurrence equation ($\ell,m
\ge 0$)
\begin{eqnarray}
\label{2drec=0} \fl
    m\,\Big[f_\eta^2\, \kappa_{\ell+4,m-1}^{(n,0)}(G,P) + (1-f_\eta^2) \,\kappa_{\ell+2,m-1}^{(n,0)}(G,P)\Big] - 4n\,
     f_\eta \,\kappa_{\ell+2,m}^{(n,0)}(G,P) \nonumber \\
    \fl \qquad\qquad\quad
    - 2\left(\ell+ 2 m + 1 \right) \, \kappa_{\ell+1,m}^{(n,0)}(G,P)    +  (2\ell+3m+2) \, \kappa_{\ell,m+1}^{(n,0)}(G,P)
    \nonumber\\
    \fl \qquad\qquad\quad
     + \, 6m \,f_\eta^2
    \sum_{i=0}^{m-1}\left( {m-1}\atop{i} \right) \sum_{j=0}^\ell   \left( {\ell}\atop{j} \right)
    \kappa_{j+2,i}^{(n,0)}(G,P)\, \kappa_{\ell-j+2,m-i-1}^{(n,0)}(G,P)=0,\nonumber\\
    {}
\end{eqnarray}
supplemented by yet another recurrence ($\ell \ge 2$), see Eq.~(\ref{cumeq}),
\begin{eqnarray}
\label{cumeq=0}
\fl (4n^2-\ell^2)\,(\ell+1) \kappa_{\ell+1,0}^{(n,0)}
        + 2n (2\ell-1)\,\ell \kappa_{\ell,0}^{(n,0)} \nonumber\\
        \fl \qquad + \ell(\ell-1)(\ell-2)\,
     \kappa_{\ell-1,0}^{(n,0)}
     - 2 \sum_{j=0}^{\ell-1} (3j+1) (j-\ell)^2\left({\ell}\atop{j}\right)
     \kappa_{j+1,0}^{(n,0)}\, \kappa_{\ell-j,0}^{(n,0)} = 0
\end{eqnarray}
that brings, in turn, a set of initial conditions $\{\kappa_{\ell,0}^{(n,0)}\}=\{\kappa_\ell^{(n,0)}(G)\}$
to Eq.~(\ref{2drec=0}).

We start an asymptotic analysis of the recurrence
Eq.~(\ref{2drec=0}) with singling out the large-$n$ Gaussian part
from $\kappa_{\ell,m}^{(n,0)}(G,P)$:
\begin{eqnarray}\fl \label{klm-gauss}
    \kappa_{\ell,m}^{(n,0)}(G,P) = \frac{n}{2} \left[
    \delta_{\ell,1}\delta_{m,0} + \left(1+\frac{f_\eta}{4}\right)\delta_{\ell,0}\delta_{m,1}
    \right]\nonumber\\
    + \frac{1}{16}
    \left[\delta_{\ell,1}\delta_{m,1} +
        \delta_{\ell,2}\delta_{m,0}  + \left(
            1+ \frac{f_\eta^2}{8}
        \right)\delta_{\ell,0}\delta_{m,2}
    \right] +\delta\kappa_{\ell,m}.
\end{eqnarray}
The Gaussian part was read off from Eqs.~(\ref{k-l-1}) and
(\ref{k-l-2}); the term $\delta\kappa_{\ell,m}$ accommodates
non-Gaussian corrections to the joint cumulants of Landauer
conductance and the noise power. Similarly to the asymptotic analysis of Landauer cumulants (Section \ref{cum-fo}), their large-$n$ behavior can be
studied by employing the $1/n$ expansion
\begin{eqnarray}
\label{ans-joint}
    \delta\kappa_{\ell,m} = \frac{1}{n^{\ell+m}}
    \left(
    \alpha_{\ell,m} + \frac{\beta_{\ell,m}}{n} + \frac{\gamma_{\ell,m}}{n^2} + {\mathcal O}(n^{-3}) \right),
\end{eqnarray}
where $\alpha_{1,0}=0$. Substitution of Eqs.~(\ref{klm-gauss})
and (\ref{ans-joint}) into the two-dimensional recurrence
Eq.~(\ref{2drec=0}) brings yet another recurrence equation
($\ell+m>0$):
\begin{eqnarray} \label{rr-a}\fl
    m\left(
        1 - \frac{f_\eta^2}{4}
    \right)\, \alpha_{\ell+2,m-1} - 4 f_\eta \alpha_{\ell+2,m} \nonumber\\
        -2 (\ell+1+2m) \alpha_{\ell+1,m} + (2\ell+2+3m)\,\alpha_{\ell,m+1}=0.
\end{eqnarray}
Its (unique) solution, subject to the boundary condition (see Eq.~(\ref{a-un-fa}) in Section \ref{cum-fo})
\begin{eqnarray}
    \alpha_{\ell,0} =\frac{a_\ell(0)}{2^{2\ell}} =\frac{(\ell-1)!}{2^{2\ell+3}} \left[
    1 + (-1)^\ell
    \right],
\end{eqnarray}
reads ($\ell+m>0$):
\begin{eqnarray}
    \alpha_{\ell,m} =  \frac{(\ell+m-1)!}{2^{2(\ell+m)+3}}
    \left[
    \left(
        \frac{f_\eta}{2}+1
    \right)^m + (-1)^\ell \left(
        \frac{f_\eta}{2}-1
    \right)^m
    \right].
\end{eqnarray}
Taken together with Eq.~(\ref{ans-joint}), it yields the leading
term in the $1/n$ expansion for joint cumulants of Landauer
conductance and the noise power:
\begin{eqnarray}
\label{ans-joint-answer}
    \delta\kappa_{\ell,m} =\frac{1}{8}
    \frac{(\ell+m-1)!}{(4n)^{\ell+m}}
    \left[
    \left(
        \frac{f_\eta}{2}+1
    \right)^m + (-1)^\ell \left(
        \frac{f_\eta}{2}-1
    \right)^m
    \right].
\end{eqnarray}
Equations (\ref{klm-gauss}) and (\ref{ans-joint-answer}) are the
central result of this Section.

In particular, they bring the following large-$n$ expression for
the cumulants of noise power in case of symmetric leads:
\begin{eqnarray} \label{npc} \fl
    \kappa_\ell^{(n,0)}(P) =
     2n \left(
    1+\frac{f_\eta}{4}
    \right) \delta_{\ell,1} + \left(
    1 + \frac{f_\eta^2}{8}
    \right) \delta_{\ell,2} \nonumber \\
    + \frac{(\ell-1)!}{8n^\ell}
    \left[
    \left( \frac{f_\eta}{2}-1
    \right)^\ell
    +
    \left( \frac{f_\eta}{2}+1
    \right)^\ell
    \right] + o(n^{-\ell}).
\end{eqnarray}
The higher order corrections to the above result can straightforwardly be found as well. An extension of the above results to $\beta=1$ and $4$ symmetry classes can be found in Mezzadri and Simm (2013).

\subsection{Brief summary}
In the previous sections, we have shown how the ideas of integrability can be utilized to provide
a nonperturbative description of universal aspects of quantum transport through chaotic cavities connected to the external world
through ideal leads. Focussing on fluctuation properties of the Landauer conductance and the noise power, we revealed that they
are remarkably (and quite unexpectedly!) described in terms of fifth Painlev\'e transcendents and the one-dimensional Toda Lattice. While the
formalism presented refers to the cavities with broken time-reversal symmetry ($\beta=2$ Dyson's symmetry class), an integrable theory of
universal quantum transport can equally be extended to other fundamental symmetry classes ($\beta=1$ and $4$) and other transport observables, see a detailed exposition by
Mezzadri and Simm (2013).

Importantly, integrable theory of universal quantum transport presented in Sections \ref{SoLC} and \ref{itnf} assumed that a chaotic cavity is probed via ideal leads attached to the cavity through ballistic point contacts (so that the ${\bm {\mathcal S}}$-matrix is CUE-distributed). Will the integrability persist if one relaxes the ballisticity assumption? Below, this question will be answered in the affirmative.

\section[Chaotic cavities with non-ideal leads and two-dimensional Toda Lattice]{Quantum transport in chaotic cavities with non-ideal leads and two-dimensional Toda Lattice}\label{Tunneling}

\subsection{Joint probability density function of reflection eigenvalues}\label{jpdf-ref-eig}

In case of non-ideal leads, the underlying assumption of Sections \ref{SoLC} and {\ref{itnf}} about uniform distribution of the scattering matrix ${\bm {\mathcal S}}$ with respect to the proper Haar measure is no longer justified. Instead, fluctuations of the scattering matrix are captured by the full Poisson kernel [Eq.~(\ref{pk})] which, in turn, induces a nontrivial probability measure for a set ${\bm T}$ of all non-zero transmission eigenvalues.

A first systematic random-matrix-theory study \footnote[1]{For an earlier RMT research in the field, see: Brouwer and Beenakker (1994, 1995).} of a largely unexplored territory of non-ideal couplings had recently been undertaken by Vidal and Kanzieper (2012) who managed to lift a point-contact-ballisticity for one (`left') of the two leads. To address integrability issues we are concerned with, we shall choose a (reduced) model of {\it mode-independent tunneling}, when the average scattering matrix equals [Eq.~(\ref{gj})]
\begin{eqnarray}\label{s0-def}
    {\bm {\mathcal S}}_0 = {\bm V}^\dagger
    \left(
                      \begin{array}{cc}
                 \hat{\bm \gamma}_{\rm L}  & 0 \\
                 0 & 0 \\
               \end{array}
                   \right)
    {\bm V},\qquad \hat{\bm \gamma}_L = \mathds{1}_{N_{\rm L}}\sqrt{1 - \Gamma^2}.
\end{eqnarray}
Here, $\Gamma = \sqrt{1-\gamma_{\rm L}^2}$ is a tunnel probability for the $j$-th propagating mode ($1\le j \le N_{\rm L}$) in the left, non-ideal lead; the right lead, supporting $N_{\rm R}\ge N_{\rm L}$ open channels, is kept ideal. The joint probability density function of transmission/reflection eigenvalues is the central object of our interest in this Section. It will be derived in three steps.
\noindent\newline\newline
{\it Step No 1}.---Spotting that, in the above setting, the Poisson kernel solely depends on the reflection matrix ${\bm r}$ in Eq.~(\ref{s-block}), and assuming broken time-reversal symmetry ($\beta=2$), one observes
\begin{eqnarray}
    P({\bm {\mathcal S}}) \propto {\rm det}^{-N} (\mathds{1}_{N_{\rm L}} - \gamma_{\rm L} {\bm r})
    \, {\rm det}^{-N} (\mathds{1}_{N_{\rm L}} - \gamma_{\rm L} {\bm r}^\dagger).
\end{eqnarray}
Consequently, the probability density associated with the reflection matrix alone equals
\begin{eqnarray}\label{pr-from-ps}\fl
 P({\bm r}) \propto  {\rm det}^{-N} (\mathds{1}_{N_{\rm L}} - \gamma_{\rm L} {\bm r})
    \, {\rm det}^{-N} (\mathds{1}_{N_{\rm L}} - \gamma_{\rm L} {\bm r}^\dagger)
    \int_{{\bm t} \in {\mathbb C}^{N_{\rm L}\times N_{\rm R}}} [d{\bm t}] \,\delta ({\bm t}{\bm t}^\dagger + {\bm r}{\bm r}^\dagger-{\mathds 1}_{N_{\rm L}}),\nonumber\\
    {}
\end{eqnarray}
where the integral over complex valued transmission matrix ${\bm t} \in {\mathbb C}^{N_{\rm L}\times N_{\rm R}}$ originates from the unitarity of the scattering matrix, ${\bm {\mathcal S}} {\bm {\mathcal S}}^\dagger = \mathds{1}_N$. Performing \footnote[4]{See, e.g., Lemma 1, Appendix A in: Kanzieper and Singh (2010).} the integral in Eq.~(\ref{pr-from-ps}),
\begin{eqnarray} \fl
    \int_{{\bm t} \in {\mathbb C}^{N_{\rm L}\times N_{\rm R}}} [d{\bm t}] \,\delta ({\bm t}{\bm t}^\dagger + {\bm r}{\bm r}^\dagger-{\mathds 1}_{N_{\rm L}}) \propto {\rm det}^{N_{\rm R}-N_{\rm L}} (\mathds{1}_{N_{\rm L}} - {\bm r} {\bm r}^\dagger) \, \Theta (\mathds{1}_{N_{\rm L}} - {\bm r} {\bm r}^\dagger),
\end{eqnarray}
we obtain:
\begin{eqnarray}\label{pr-int}
 P({\bm r}) \propto  {\rm det}^{-N} (\mathds{1}_{N_{\rm L}} - \gamma_{\rm L} {\bm r})
    \, {\rm det}^{-N} (\mathds{1}_{N_{\rm L}} - \gamma_{\rm L} {\bm r}^\dagger)\, \nonumber\\
    \qquad \qquad \times  {\rm det}^{N_{\rm R}-N_{\rm L}} (\mathds{1}_{N_{\rm L}} - {\bm r} {\bm r}^\dagger) \, \Theta (\mathds{1}_{N_{\rm L}} - {\bm r} {\bm r}^\dagger).
\end{eqnarray}
\noindent\newline
{\it Step No 2.}---Since ${\bm r}$ is a complex valued matrix with no symmetries, it can be pseudo-diagonalized via singular value decomposition, ${\bm r} = {\bm u} {\bm \varrho} {\bm v}^\dagger$, where ${\bm \varrho} = {\rm diag}(\varrho_1,\dots,\varrho_{N_{\rm L}})$ with $\varrho_j>0$ and the unitary matrices ${\bm u}$ and ${\bm v}$ are ${\bm u}\in U(N_{\rm L})$ and ${\bm v}\in U(N_{\rm L})/U(1)^{N_{\rm L}}$. Further, introducing reflection eigenvalues $R_j=\rho_j^2$ (these are the eigenvalues of the matrix ${\bm r}{\bm r}^\dagger$), and observing the relation (see, e.g., Forrester 2006)
\begin{eqnarray}
    [d{\bm r}] \propto \Delta_{N_{\rm L}}^2({\bm R})\, d\mu({\bm u}) \, d\mu({\bm v})\,\prod_{j=1}^{N_{\rm L}} dR_j
\end{eqnarray}
(here, $d\mu$ is the invariant Haar measure on the unitary group), we come down to the joint probability density of reflection eigenvalues in the form:
\begin{eqnarray}\fl \label{PR-01}
    P(R_1,\dots,R_{N_{\rm L}};\gamma^2) &\propto& \Delta_{N_{\rm L}}^2({\bm R}) \, \prod_{j=1}^{N_{\rm L}} (1-R_j)^{N_{\rm R}-N_{\rm L}} \Theta(1-R_j)\, dR_j \nonumber\\
    &\times& \int_{{\bm U}\in U(N_{\rm L})} d\mu({\bm U})\,
    {\rm det}^{-N} (\mathds{1}_{N_{\rm L}} - \gamma_{\rm L} {\bm \varrho}{\bm U})\,
    {\rm det}^{-N} (\mathds{1}_{N_{\rm L}} - \gamma_{\rm L} {\bm \varrho} {\bm U}^\dagger).\nonumber\\
    {}
\end{eqnarray}
Notice, that for any finite $\gamma_{\rm L}$, the $U(N_{\rm L})$ group integral in Eq.~(\ref{PR-01}) effectively modifies the interaction between reflection eigenvalues, which is no longer logarithmic [see Eq.~(\ref{PnT})].
\noindent\newline\newline
{\it Step No 3.}---The group integral in Eq.~(\ref{PR-01}) can be evaluated by employing the technique of Schur functions (Macdonald 1995) and the theory of hypergeometric functions of matrix argument (Muirhead 2005, Gross and Richards 1989). (An alternative derivation, based on the theory of $\tau$ functions of matrix argument, was reported by Orlov (2004).) Leaving details of our calculation for a separate publication, we state the final result:
\begin{eqnarray} \label{final-dgi}\fl
\int_{{\bm U}\in U(N_{\rm L})} d\mu({\bm U})\,
    {\rm det}^{-N} (\mathds{1}_{N_{\rm L}} - \gamma_{\rm L} {\bm \varrho}{\bm U})\,
    {\rm det}^{-N} (\mathds{1}_{N_{\rm L}} - \gamma_{\rm L} {\bm \varrho} {\bm U}^\dagger)
    \nonumber\\
    = \frac{1}{\Delta_{N_{\rm L}}({\bm R})}\, {\rm det}_{(j,k) \in (1, N_{\rm L})}
    \left[
        R_j^{k-1} {}_2 F_{1} (N_{\rm R}+k, N_{\rm R}+k; k; \gamma^2 R_j)
    \right].
\end{eqnarray}
Here, ${}_2 F_1$ is the Gauss hypergeometric function.
\noindent\newline\newline
{\it Step No 4.}---Combining the last two equations together, we eventually derive the sought joint probability density function of reflection eigenvalues:
\begin{eqnarray}\fl \label{PR-02}
    P(R_1,\dots,R_{N_{\rm L}};\gamma^2) = c(N_{\rm L}, N_{\rm R}) \, (1-\gamma^2)^{N_{\rm L}(N_{\rm L}+N_{\rm R})} \Delta_{N_{\rm L}}({\bm R}) \, \prod_{j=1}^{N_{\rm L}} (1-R_j)^{N_{\rm R}-N_{\rm L}} \nonumber\\
    \times
    \, {\rm det}_{(j,k) \in (1, N_{\rm L})}
    \left[
        R_j^{k-1} {}_2 F_{1} (N_{\rm R}+k, N_{\rm R}+k; k; \gamma^2 R_j)
    \right].
\end{eqnarray}
Here, we reinstated the factor $(1-\gamma^2)^{N_{\rm L}(N_{\rm L}+N_{\rm R})}$ originating from the normalization of the Poisson kernel
$\propto {\rm det}^{-N}(\mathds{1}_{N} - {\bm {\mathcal S}}_0 {\bm {\mathcal S}}_0^\dagger)$; the otherwise $\gamma$-independent prefactor
$c(N_{\rm L}, N_{\rm R})$ can easily be restored from the $\gamma\rightarrow 0$ limit, where the corresponding joint probability density function of reflection/transmission eigenvalues is known [Eqs.~(\ref{PnT}) and (\ref{nc})]. This leads to
\begin{eqnarray}
    c(N_{\rm L},N_{\rm R}) = \frac{N_{\rm L}! \, N_{\rm R}!}{(N_{\rm L}+N_{\rm R})!}
    \prod_{j=1}^{N_{\rm L}} \frac{1}{(j!)^2} \prod_{j=1}^{N_{\rm L}} \frac{(N_{\rm R}+j)!}{(N_{\rm R}-j)!}.
\end{eqnarray}
This completes our derivation of the joint probability density of reflection eigenvalues in a chaotic cavity probed via one ideal and one non-ideal lead. For a more general case of mode-dependent tunneling (when $\hat{\bm \gamma}_{\rm L}$ in Eq.~(\ref{s0-def}) is not proportional to the unity matrix ${\mathds 1}_{N_{\rm L}}$), the reader is referred to Vidal and Kanzieper (2012).

\subsection{Conductance fluctuations and two-dimensional Toda Lattice}
To quantify statistics of the Landauer conductance for a cavity probed via both ballistic (right lead) and tunnel (left lead) point contacts, we concentrate on the MGF
\begin{eqnarray}\fl
    {\mathcal F}^{(N_{\rm L}, N_{\rm R})} (\gamma^2, z) = \int_{(0,1)^{N_{\rm L}}} \prod_{j=1}^{N_{\rm L}}\left( dR_j \, e^{-z(1-R_j)} \right)\,
    P(R_1,\dots,R_{N_{\rm L}};\gamma^2),
\end{eqnarray}
where the joint probability density function in the integrand was determined in the previous Section. Having in mind Eq.~(\ref{PR-02}), one has:
\begin{eqnarray}\fl
    {\mathcal F}^{(N_{\rm L}, N_{\rm R})} (\gamma^2, z) = c_\gamma(N_{\rm L}, N_{\rm R})\, e^{-N_{\rm L} z} \int_{(0,1)^{N_{\rm L}}} \prod_{j=1}^{N_{\rm L}}
    \left( dR_j \, e^{z R_j} (1-R_j)^{N_{\rm R}-N_{\rm L}}\right)\,\nonumber\\
    \times \Delta_{N_{\rm L}}({\bm R})\,
    {\rm det}_{(j,k) \in (1, N_{\rm L})}
    \left[
        R_j^{k-1} {}_2 F_{1} (N_{\rm R}+k, N_{\rm R}+k; k; \gamma^2 R_j)
    \right],
\end{eqnarray}
where
\begin{eqnarray}
    c_\gamma(N_{\rm L}, N_{\rm R})=c(N_{\rm L}, N_{\rm R}) \, (1-\gamma^2)^{N_{\rm L}(N_{\rm L}+N_{\rm R})}.
\end{eqnarray}
Owing to the Andr\'eief--de Bruijn integration formula, this can further be reduced down to
\begin{eqnarray}\label{mgf-tun-01}\fl
    {\mathcal F}^{(N_{\rm L}, N_{\rm R})} (\gamma^2, z) = N_{\rm L}!\, c_\gamma(N_{\rm L}, N_{\rm R})\, e^{-N_{\rm L} z}
    \, {\rm det}_{(j,k) \in (1, N_{\rm L})}
    \left[ {\mathcal M}_{jk}^{(N_{\rm L},N_{\rm R})}(\gamma^2,z)
    \right]
\end{eqnarray}
with
\begin{eqnarray}\fl
    {\mathcal M}_{jk}^{(N_{\rm L},N_{\rm R})}(\gamma^2,z) =
        \int_0^1 dR\, e^{z R} (1-R)^{N_{\rm R}-N_{\rm L}}\, R^{j+k-2}\, {}_2 F_{1} (N_{\rm R}+k, N_{\rm R}+k; k; \gamma^2 R)
         \nonumber\\
        \fl\qquad
        = \left(\frac{\partial}{\partial z}\right)^{j-1}
        \,\int_0^1 dR\, e^{z R} (1-R)^{N_{\rm R}-N_{\rm L}}\, R^{k-1}\, {}_2 F_{1} (N_{\rm R}+k, N_{\rm R}+k; k; \gamma^2 R).
\end{eqnarray}
The entry ${\mathcal M}_{jk}^{(N_{\rm L},N_{\rm R})}(\gamma^2,z)$ can be written in a more appealing form after one makes use of the identity
\begin{eqnarray}\fl
        R^{k-1}\, {}_2 F_{1} (N_{\rm R}+k, N_{\rm R}+k; k; \gamma^2 R) =(k-1)!\nonumber\\
        \times  \left(\frac{N_{\rm R}!}{(N_{\rm R}+k-1)!} \right)^2
        \left( \frac{\partial}{\partial \gamma^2}\right)^{k-1} {}_2 F_1 (N_{\rm R}+1, N_{\rm R}+1; 1; \gamma^2 R)
\end{eqnarray}
that yields the representation
\begin{eqnarray}\fl \label{mjk}
    {\mathcal M}_{jk}^{(N_{\rm L},N_{\rm R})}(\gamma^2,z) = (k-1)! \left(\frac{N_{\rm R}!}{(N_{\rm R}+k-1)!} \right)^2
         \left(\frac{\partial}{\partial z}\right)^{j-1}
         \left( \frac{\partial}{\partial \gamma^2}\right)^{k-1}
        {\mathcal M}_{11}^{(N_{\rm L},N_{\rm R})}(\gamma^2,z). \nonumber\\
        {}
\end{eqnarray}
Here,
\begin{eqnarray} \fl
    {\mathcal M}_{11}^{(N_{\rm L},N_{\rm R})}(\gamma^2,z)=\int_0^1 dR\, e^{z R} (1-R)^{N_{\rm R}-N_{\rm L}}\,
            {}_2 F_{1} (N_{\rm R}+1, N_{\rm R}+1; 1; \gamma^2 R).
\end{eqnarray}
With the help of Eqs.~(\ref{mgf-tun-01}) and (\ref{mjk}), we may now put the MGF into a fairly symmetric form:
\begin{eqnarray}\label{mgf-tun-02}\fl
    {\mathcal F}^{(N_{\rm L}, N_{\rm R})} (\gamma^2, z) = \tilde{c}_\gamma(N_{\rm L}, N_{\rm R})\, e^{-N_{\rm L} z} \nonumber\\
    \times
    \, {\rm det}_{(j,k) \in (1, N_{\rm L})}
    \left[
    \left( \frac{\partial}{\partial z}\right)^{j-1} \left( \frac{\partial}{\partial \gamma^2}\right)^{k-1}
    {\mathcal M}_{11}^{(N_{\rm L},N_{\rm R})}(\gamma^2,z) \right],
\end{eqnarray}
where
\begin{eqnarray}\fl
    \tilde{c}_\gamma(N_{\rm L},N_{\rm R}) = N_{\rm L}! \times (1-\gamma^2)^{N_{\rm L}(N_{\rm L}+N_{\rm R})}
    \,\frac{(N_{\rm L}+N_{\rm R})!}{N_{\rm L}! N_{\rm R}!}
    \prod_{j=0}^{N_{\rm L}-1} \frac{1}{j!} \prod_{j=1}^{N_{\rm L}} \frac{(N_{\rm R}!)^2}{(N_{\rm R}-j)!\,(N_{\rm R}+j)!}.
    \nonumber\\
    {}
\end{eqnarray}

Equation (\ref{mgf-tun-02}) suggests that the conductance MGF ${\mathcal F}_\gamma(N_{\rm L}, N_{\rm R}; z)$ considered as a function of two continuous variables $(z,\gamma^2)$ can be related to a solution of a {\it two-dimensional Toda Lattice} equation. To make this statement precise, we need the theorem (see, e.g., Section 6 in: Vein and Dale 1999) going back to Darboux (1889) and stating that for a differentiable function $f(x,y)$ the determinant
\begin{eqnarray}
    u_n(x,y) = {\rm det}_{(j,k)\in (1,n)} \left[
    \left(\frac{\partial}{\partial x}\right)^{j-1} \left(\frac{\partial}{\partial y}\right)^{k-1} f(x,y)
    \right]
\end{eqnarray}
satisfies the two-dimensional Toda Lattice equation:
\begin{eqnarray}
    \frac{\partial^2 }{\partial x \partial y} \log u_n(x,y) = \frac{u_{n-1} \, u_{n+1}}{u_n^2}.
\end{eqnarray}
It is assumed that $u_0(x,y)=1$.

With the above in mind, we are ready to make a central statement of this Section. Let $\{u_n^{(N_{\rm L}, N_{\rm R})}(\gamma^2, z)\}$ denotes a sequence of determinants
\begin{eqnarray}\fl
    u_n^{(N_{\rm L}, N_{\rm R})}(\gamma^2, z) = {\rm det}_{(j,k)\in (1,n)} \left[
    \left(\frac{\partial}{\partial z}\right)^{j-1} \left(\frac{\partial}{\partial \gamma^2}\right)^{k-1} {\mathcal M}_{11}^{(N_{\rm L},N_{\rm R})}(\gamma^2,z)
    \right]
\end{eqnarray}
for $n=1,2,\dots$. Then, $\{u_n^{(N_{\rm L}, N_{\rm R})}(\gamma^2, z)\}$ satisfies the two-dimensional Toda Lattice equation
\begin{eqnarray}
    \frac{\partial^2 }{\partial z \partial \gamma^2} \log u_n^{(N_{\rm L}, N_{\rm R})}(\gamma^2, z) =
    \frac{u_{n-1}^{(N_{\rm L}, N_{\rm R})}(\gamma^2, z)\, u_{n+1}^{(N_{\rm L}, N_{\rm R})}(\gamma^2, z)} {\left(u_{n}^{(N_{\rm L}, N_{\rm R})}
    (\gamma^2, z)\right)^2},
\end{eqnarray}
and the MGF for Landauer conductance equals the $N_{\rm L}$-th member of the above sequence:
\begin{eqnarray}
    {\mathcal F}^{(N_{\rm L}, N_{\rm R})} (\gamma^2, z) = u_{n}^{(N_{\rm L}, N_{\rm R})}(\gamma^2, z)\Big|_{n=N_{\rm L}}.
\end{eqnarray}
This concludes our derivation of the announced relation between the problem of conductance fluctuations in chaotic cavities with a non-ideal lead and the two-dimensional Toda Lattice equation.

\section{Concluding remarks}
In this paper, we reviewed the basics of integrable theory of the universal quantum transport in chaotic structures. (i) Starting with the mathematically simplest case of chaotic cavities with broken time reversal symmetry ($\beta=2$) which are coupled to the outside world through ballistic point contacts (`ideal leads'), we showed that fluctuations of the Landauer conductance and the noise power are described by a {\it one-dimensional Toda Lattice} equation and the {\it fifth Painl\'eve transcendent}. This finding, revealed at the level of moment generating functions, was utilized to generate nonlinear recurrence relations between cumulants of transport observables of various orders. Solutions to these relations produced non-perturbative formulae for cumulants of the Landauer conductance and the noise power of any given order. The discovered relation between quantum transport in zero-dimensional chaotic cavities and the theory of integrable lattices appears to be very general and can be carried over to scattering systems with preserved time-reversal symmetry ($\beta=1$ and $4$); there, fluctuations of transport observables are governed by a {\it Pfaff-Toda Lattice} (Mezzadri and Simm 2013).

(ii) Further, we demonstrated that inclusion of tunneling effects inherent in realistic point contacts (`non-ideal leads') does not destroy the integrability. Specifically, for $\beta=2$ symmetry class, fluctuations of the Landauer conductance for a cavity probed via both ideal and non-ideal leads were shown to be captured by a {\it two-dimensional Toda Lattice} equation. While this result marks quite a progress in understanding integrable aspects of the universal quantum transport, more efforts are required to bring an integrable theory of quantum transport to its culminating point: (i) extending the formalism to other Altland-Zirnbauer symmetry classes (Altland and Zirnbauer 1997) and, even more important, (ii) relaxing a point contact ballisticity for the second lead are the most challenging problems whose solution is very much called for. A progress in solving the former problem will be reported elsewhere (Jarosz, Vidal and Kanzieper 2014).

\section*{Acknowledgements}
I am indebted to A.~Jarosz, V.~Al.~Osipov and P.~Vidal for collaboration on the
integrability project that led, in part, to the results reported in this contribution. This work was supported by the Israel Science Foundation through the grants No 414/08 and No 647/12.
\smallskip\smallskip\smallskip

\newpage

\section*{References}
\fancyhead{} \fancyhead[RE,LO]{\small{References}}
\fancyhead[LE,RO]{\thepage}

\begin{harvard}

\item[] Adler M and van Moerbeke P 1995
        Matrix integrals, Toda symmetries, Virasoro constraints, and orthogonal polynomials
        {\it Duke Math J} {\bf 80} 863

\item[] Adler M, Shiota T and van Moerbeke P 1995
        Random matrices, vertex operators and the Virasoro algebra
        {\it Phys. Lett. A} {\bf 208} 67

\item[] Adler M and van Moerbeke P 2001
        Hermitian, symmetric and symplectic random ensembles: PDE's for the distribution of the spectrum
        {\it Ann. Math.} {\bf 153}, 149

\item[] Agam O, Aleiner I and Larkin A 2000
        Shot noise in chaotic systems: "Classical" to quantum crossover
        {\it Phys Rev Lett} {\bf 85} 3153

\item[] Adagideli \.{I} 2003
        Ehrenfest-time-dependent suppression of weak localization
        {\it Phys Rev B} {\bf 68} 233308

\item[] Aleiner I L and Larkin A I 1996
        Divergence of classical trajectories and weak localization
        {\it Phys Rev B} {\bf 54} 14423

\item[] Aleiner I L and Larkin A I 1997
        Role of divergence of classical trajectories in quantum chaos
        {\it Phys Rev E} {\bf 55} R1243

\item[] Alhassid Y 2000
        The statistical theory of quantum dots
        {\it Rev Mod Phys} {\bf 72} 895

\item[] Altland A and Zirnbauer M 1997
        Non-standard symmetry classes in mesoscopic normal-/superconducting hybrid structures
        {\it Phys Rev B} {\bf 55} 1142

\item[] Andr\'eief C 1883
        Note sur une relation les int\'egrales d\'efinies des produits des fonctions
        {\it M\'em de
        la Soc Sci} {\bf 2} 1

\item[] Baranger H U and Mello P A 1994
        Mesoscopic transport through chaotic cavities: A random S-matrix theory approach
        {\it Phys Rev Lett} {\bf 73} 142

\item[] Beenakker C W J 1997
        Random matrix theory of quantum transport
        {\it Rev. Mod. Phys.} {\bf 69}, 731

\item[] Blanter Ya M and B\"uttiker M 2000
        Shot noise in mesoscopic conductors
        {\it Phys Rep} {\bf 336} 1

\item[] Blanter Ya M and Sukhorukov E V 2000
        Semiclassical theory of conductance and noise in open chaotic cavities
        {\it Phys Rev Lett} {\bf 84} 1280

\item[] Bl\"umel R and Smilansky U 1990
       Random-matrix description of chaotic scattering: Semiclassical approach
       {\it Phys. Rev. Lett.} {\bf 64}, 241

\item[] Bohigas O, Giannoni M-J and Schmit C 1984
        Characterization of chaotic quantum spectra and universality of level fluctuation laws
        {\it Phys Rev Lett} {\bf 52} 1

\item[] Braun P, Heusler S, M\"uller S and Haake F 2006
        Semiclassical prediction for shot noise in chaotic cavities
        {\it J Phys A: Math Gen} {\bf 39} L159

\item[] Brouwer P W and Beenakker C W J 1994
        Conductance distribution of a quantum dot with nonideal single-channel leads
        {\it Phys Rev B} {\bf 50} R11263

\item[] Brouwer P W 1995
        Generalized circular ensemble of scattering matrices for a chaotic cavity with non-ideal leads
        {\it Phys Rev B} {\bf 51} 16878

\item[] Brouwer P W and Beenakker C W J 1995
        Effect of a voltage probe on the phase-coherent conductance of a ballistic chaotic cavity
        {\it Phys Rev B} {\bf 51} 7739

\item[] Brouwer P W and Beenakker C W J 1996
        Diagrammatic method of integration over the unitary group, with applications to quantum transport in mesoscopic systems
        {\it J Math Phys} {\bf 37} 4904

\item[] Brouwer P W and Rahav S 2006
        Semiclassical theory of the Ehrenfest time dependence of quantum transport in ballistic quantum dots
        {\it Phys Rev B} {\bf 74} 075322

\item[] Brouwer P W 2007
        Semiclassical theory of the Ehrenfest-time dependence of quantum transport
        {\it Phys Rev B} {\bf 76} 165313

\item[] B\"uttiker M 1990
        Scattering theory of thermal and excess noise in open conductors
        {\it Phys~Rev~Lett} {\bf 65} 2901

\item[] B\"uttiker M 1992
        Scattering theory of current and intensity noise correlations in conductors and wave guides
        {\it Phys~Rev~B} {\bf 46} 12485

\item[] Chazy J 1911
        Sur les \'equations diff\'erentielles du troisi\`{e}me ordre et
        d'ordre sup\'erieur dont l'int\'egrale g\'en\'erale a ses
        points critiques fixes
        {\it Acta Math} {\bf 34} 317

\item[] Clarkson P A 2003
        Painlev\'e equations -- nonlinear special functions
        {\it J Comp Appl Math} {\bf 153} 127

\item[] Cosgrove C M and Scoufis G 1993
        Painlev\'e classification of a class of differential equations
        of the second order and second degree
        {\it Stud Appl Math} {\bf 88} 25

\item[] Cron R, Goffman M F, Esteve D and Urbina C 2001
        Multiple-charge-quanta shot noise in superconducting atomic contacts
        {\it Phys Rev Lett} {\bf 86} 4104

\item[] Darboux G 1889
        {\it Le\c{c}ons sur la Th\'{e}orie G\'{e}n\'{e}rale des Surfaces et les Applications G\'{e}om\'{e}triques du
        Calcul Infinit\'{e}simal}. Deuxi\'{e}me Partie (Gauthier-Villars Et Fils, Paris)

\item[] Date E, Kashiwara M, Jimbo M and Miwa T 1983 in:
        {\it Nonlinear Integrable Systems—Classical Theory and
        Quantum Theory} edited by Jimbo M and Miwa T
        (World Scientific: Singapore)

\item[]  de Bruijn N G 1955
         On some multiple integrals involving determinants
         {\it J Indian Math Soc} {\bf 19} 133

\item[] Efetov K 1997
        {\it Supersymmetry in Disorder and Chaos}
        (Cambridge University Press, Cambridge)

\item[] Fisher D S and Lee P 1981
        Relation between conductivity and transmission matrix
        {\it Phys Rev B} {\bf 23} R6851

\item[] Forrester P J and Witte N S 2002
        Application of the $\tau$ function theory of Painlev\'e
        equations to random matrices: PV, PIII, the LUE, JUE and CUE
        {\it Commun Pure Appl Math} {\bf 55} 679

\item[] Forrester P J 2006
        Quantum conductance problems and the Jacobi ensemble
        {\it J Phys A: Math Gen.} {\bf 39} 6861

\item[] Forrester P J 2010
        {\it Log-Gases and Random Matrices}
        (Princeton University Press, Princeton)

\item[] Gangardt D M and Kamenev A
        Replica treatment of the Calogero-Sutherland model
        {\it Nucl Phys B} {\bf 610} 575

\item[] Gross K I and Richards D St P 1989
        Total positivity, spherical series, and hypergeometric functions of matrix argument
        {\it J Approx Th} {\bf 59} 224

\item[] Harish-Chandra 1957
        Differential operators on a semisimple Lie algebra
        {\it Amer J Math} {\bf 79} 87

\item[] Heusler S, M\"uller S, Braun P and Haake F 2006
        Semiclassical theory of chaotic conductors
        {\it Phys Rev Lett} {\bf 96} 066804

\item[] Hua L K 1963
        {\it Harmonic Analysis of Functions of Several Complex Variables in the Classical Domains}
        (American Mathematical Society, Providence)

\item[] Imry Y 2002
        {\it Introduction to Mesoscopic Physics} (Oxford University Press, New York)

\item[] Itzykson C and Zuber J-B 1980
        The planar approximation. II
        {\it J Math Phys} {\bf 21} 411

\item[] Jalabert R A, Pichard J-L and Beenakker C W J 1994
        Universal quantum signatures of chaos in ballistic transport
        {\it Europhys Lett} {\bf 27} 255

\item[] Jarosz A, Vidal P and Kanzieper E 2014
        Integrable theory of quantum transport in chaotic cavities with a non-ideal lead
        at $\beta=1$ and $4$
        {\it (in preparation)}

\item[] Jimbo M, Miwa T, M\^ori Y and Sato M 1980
        Density matrix of an impenetrable Bose gas and the fifth
        Painlev\'e transcendent
        {\it Physica D} {\bf 1} 80

\item[] Kanzieper E 2002
        Replica field theories, Painlev\'e transcendents, and exact correlation functions
        {\it Phys Rev Lett} {\bf 89} 250201

\item[] Kanzieper E and Singh N 2010
        Non-Hermitean Wishart random matrices (I)
        {\it J Math Phys} {\bf 51} 103510

\item[] Khlus V A 1987
        Current and voltage fluctuations in micro-junctions of normal and superconducting metals
        {\it Zh~\'Eksp~Teor~Fiz} {\bf 93} 2179 [{\it Sov~Phys~JETP} {\bf 66} 1243]

\item[] Khoruzhenko B A, Savin D V and Sommers H-J 2009
        Systematic approach to statistics of conductance and shot-noise in chaotic cavities
        {\it Phys Rev B} {\bf 80} 125301

\item[] Landauer R 1957
        Spatial variation of currents and fields due to localized scatterers in metallic conduction
        {\it J Res Dev} {\bf 1} 223

\item[] Lesovik G B 1989
        Excess quantum noise in 2D ballistic point contacts
        {\it Pis'ma~Zh~\'Eksp~Teor~Fiz} {\bf 49} 513 [{\it JETP Lett} {\bf 49} 592]

\item[] Lesovik G B and Sadovskyy I A
        Scattering matrix approach to the description of quantum electron transport
        {\it Physics -- Uspekhi} {\bf 54} 1007 (2011)

\item[] Lewenkopf C H and Weidenm\"uller H A 1991
        Stochastic versus semiclassical approach to quantum chaotic scattering
        {\it Ann Phys} {\bf 212} 53

\item[] Martin T and Landauer R 1992
        Wave-packet approach to noise in multichannel mesoscopic systems
        {\it Phys~Rev~B} {\bf 45} 1742

\item[] Macdonald I G 1995
        {\it Symmetric Functions and Hall Polynomials} (Clarendon Press, Oxford)

\item[] Mehta M L 2004
        {\it Random Matrices} (Amsterdam, Elsevier)

\item[] Mello P A and Baranger H U 1999
        Interference phenomena in electronic transport through chaotic cavities: An information-theoretic approach
        {\it Waves Random Media} {\bf 9}, 105

\item[] Mezzadri F and Simm N J 2011
        Moments of the transmission eigenvalues, proper delay times and random matrix theory, I
        {\it J Math Phys} {\bf 52} 103511

\item[] Mezzadri F and Simm N J 2013
        $\tau$ function theory of quantum chaotic transport with $\beta=1,\, 2, \,4$
        {\it Commun Math Phys} {\bf 324} 465

\item[] Mironov A and Morozov A 1990
        On the origin of Virasoro constraints in matrix models: Lagrangian approach
        {\it Phys. Lett. B} {\bf 252} 47

\item[] Morozov A Yu 1994
        Integrability and matrix models
        {\it Uspekhi Fiz Nauk} {\bf 164} 3; {\it Phys-Usp} {\bf 37} 1

\item[] Muirhead R J 2005
        {\it Aspects of Multivariate Statistical Analysis} (Wiley, New Jersey)

\item[] M\"uller S, Heusler S, Braun P and Haake F 2007
        Semiclassical approach to chaotic quantum transport
        {\it New J Phys} {\bf 9} 12

\item[] Noumi M 2004
        {\it Painlev\'e Equations Through Symmetry} (AMS, Providence)

\item[] Novaes M 2008
        Statistics of quantum transport in chaotic cavities with
        broken time-reversal symmetry
        {\it Phys Rev B} {\bf 78} 035337

\item[] Oberholzer S, Sukhorukov E V, Strunk C, Sch\"onenberger C, Heinzel T and Holland M 2001
        Shot noise by quantum scattering in chaotic cavities
        {\it Phys Rev Lett} {\bf 86} 2114

\item[] Oberholzer S, Sukhorukov E V and Sch\"onenberger C 2002
        Crossover between classical and quantum shot noise in chaotic cavities
        {\it Nature} {\bf 415} 765

\item[] Okamoto K 1987
        Studies on the Painlev\'e equations, II. Fifth Painlev\'e equation PV
        {\it Jpn J Math} {\bf 13} 47

\item[] Orlov A Yu 2004
        New solvable matrix integrals
        {\it Int J Mod Phys A} {\bf 19} 276

\item[] Osipov V Al and Kanzieper E 2008
        Integrable theory of quantum transport in chaotic cavities
        {\it Phys Rev Lett} {\bf 101} 176804

\item[] Osipov V Al and Kanzieper E 2009
        Statistics of thermal to shot noise crossover in chaotic cavities
        {\it J Phys A: Math Theor} {\bf 42} 475101

\item[] Osipov V Al and Kanzieper E 2010
        Correlations of RMT characteristic polynomials and integrability: Hermitean matrices
        {\it Ann Phys} {\bf 325} 2251

\item[] Richter K 2000
        {\it Semiclassical Theory of Mesoscopic Quantum Systems} (Springer)

\item[] Richter K and Sieber M 2002
        Semiclassical theory of chaotic quantum transport
        {\it Phys Rev Lett} {\bf 89} 206801

\item[] Savin D V and Sommers H J 2006
        Shot noise in chaotic cavities with an arbitrary number of open channels
        {\it Phys Rev B} {\bf 73} R081307

\item[] Savin D V, Sommers H-J and Wieczorek W 2008
        Nonlinear statistics of quantum transport in chaotic cavities
        {\it Phys Rev B} {\bf 77} 125332

\item[] Teschl G 2000
        {\it Jacobi Operators and Completely Integrable Nonlinear Lattices} (American Math Soc, Providence)

\item[] Toda M 1989
        {\it Theory of Nonlinear Lattices} (Springer, Berlin)

\item[] Tracy C A and Widom H 1994
        Fredholm determinants, differential equations and matrix
        models
        {\it Commun Math Phys} {\bf 163} 33

\item[] Vein R and Dale P 1999
        {\it Determinants and Their Applications in Mathematical Physics} (Springer, New York)

\item[] Vidal P and Kanzieper E 2012
        Statistics of reflection eigenvalues in chaotic cavities with nonideal leads
        {\it Phys Rev Lett} {\bf 108} 206806

\item[] Whitney R S and Jacquod P 2006
        Shot noise in semiclassical chaotic cavities
        {\it Phys Rev Lett} {\bf 96} 206804

\smallskip
\end{harvard}

\end{document}